\documentclass[12pt,fleqn]{article}
\usepackage{times}
\usepackage{graphics,latexsym,psfrag,subfig,mathtools}
\usepackage{amsmath}
\usepackage{amssymb}
\usepackage{rotating}
\usepackage{lscape}
\usepackage{color}
\usepackage{enumerate}
\usepackage{hyperref}
\hypersetup{colorlinks = true, linktocpage = true, bookmarksopen = true, linkcolor = blue, urlcolor=blue, citecolor = blue, allcolors=blue}
\usepackage{authblk}
\usepackage[size=footnotesize]{caption}

\newcommand{\ba}{\begin{array}}
\newcommand{\ea}{\end{array}}

\usepackage{lipsum}

\begin{document}
\newcommand{\be}{\begin{equation}}
\newcommand{\ee}{\end{equation}}
\newcommand{\bc}{\begin{center}}
\newcommand{\ec}{\end{center}}
\newcommand{\bdm}{\begin{displaymath}}
\newcommand{\edm}{\end{displaymath}}
\newcommand{\ds}{\displaystyle}
\newcommand{\p}{\partial}
\newcommand{\INT}{\int\limits}
\newcommand{\SUM}{\sum\limits}
\newcommand{\bfm}[1]{\mbox{\boldmath $ #1 $}}
\newcommand{\change}[1]{{\color{red}{#1}}}
\newcommand{\note}[1]{{\color{blue}{#1}}}

\title{ \bf Modelling mass diffusion for a multi-layer \\
sphere immersed in a semi-infinite medium: \\
application to drug delivery }
\date{\normalsize }

\author[1]{Elliot~J.~Carr\footnote{Corresponding author: \href{mailto:elliot.carr@qut.edu.au}{elliot.carr@qut.edu.au}.}}
\author[2]{Giuseppe~Pontrelli}
\affil[1]{{\footnotesize School of Mathematical Sciences, Queensland University of Technology (QUT), Brisbane, Australia}}
\affil[2]{{\footnotesize Istituto per le Applicazioni del Calcolo -- CNR
Via dei Taurini 19 -- 00185 Rome, Italy}}

\maketitle

\begin{abstract}
We present a general mechanistic model of mass diffusion for a composite sphere placed in a large ambient medium. The multi-layer problem is described by a system of diffusion equations coupled via interlayer boundary conditions such as those imposing a finite mass resistance at the external surface of the sphere. While the work is applicable to the generic problem of heat or mass transfer in a multi-layer sphere, the analysis and results are presented in the context of drug kinetics for  desorbing and absorbing spherical microcapsules. We derive an analytical solution for the concentration in the sphere and in the surrounding  medium that avoids any artificial truncation at a finite distance. The closed-form solution  in each concentric layer is expressed in terms of a suitably-defined inverse Laplace transform that can be evaluated numerically. Concentration profiles and drug mass curves in the spherical layers and in the external environment are presented and the dependency of the solution on the mass transfer coefficient at the surface of the sphere analyzed. 
\end{abstract}

\vspace*{2ex}\noindent\textit{\bf Keywords}: mass diffusion; drug release; composite spheres; semi-analytical solution; Laplace transform.

\section{Introduction}
Models of mass transfer from spheres are commonly used from both a 
theoretical and applicative point of view. For example, studies on drug delivery from microsphere-shaped capsules or from lipid vesicles as liposomes are currently experiencing a growing interest in regards to the role played by the carrier's geometry, in its loading, stability, toxicity and, ultimately, release performance \cite{lar}. The purpose of these systems is to maintain a desired drug concentration in the blood or in the tissue for as long as possible. Among other concurrent effects, such as dissolution and possible degradation, diffusion remains the most important mechanism used to control the release rate from drug delivery systems \cite{siep,tim}. \par
On the other hand, encapsulation with multiple durable concentric layers enhances the mechanical stability and biocompatibility, protecting the sphere from the external environment and premature degradation. For some specific applications, a thin coating film is required to envelop the whole spherical structure to protect it from chemical aggression and mechanical erosion \cite{hen}. The applicative goal is to accurately predict the drug release profile from a spherical capsule and improve the overall therapeutic efficacy and safety of these drug carrier systems. \par 

Diffusion-driven mass transfer is normally described by Fick's first and second laws \cite{cra} and transport in porous media are governed by mass diffusion and convective flow models such as Darcy and the
Brinkman models \cite{vaf}. So far, several exact
and approximate solutions have been developed to analyze the kinetics of a dispersed solute from a polymeric matrix having a spherical shape. 
The mass diffusion problem is analogous to the problem of heat transfer from a sphere that has been solved by many authors in the past with a large amount of published models and approaches. A number of configurations of heat diffusion have been treated in the pioneering work of Carslaw and Jaeger \cite{caj}: the spherical matrix can have different surface conditions, with a prescribed inward or outward flux or be in contact with a well-stirred medium. The case of a composite sphere with no contact resistance has been solved with Laplace transform in analogy with that of contiguous slabs. Since
the 60s, Higuchi \cite{hig} derived analytic solutions for a single sphere in a perfect sink using pseudo-state approximations, without a boundary layer effect. In the classical book of Crank \cite{cra}, the diffusion in a sphere from a well-stirred medium is solved by Fourier expansion in the case of constant or time varying surface concentration and the case of a constant flux at the surface. 
More recent work includes empirical, semi-empirical and
mechanistic diffusion models \cite{cuo}. An exact solution for diffusional release of a dispersed solute from a spherical polymer matrix into both semi-infinite and finite external mediums has been developed \cite{abd,zhou}. Simulations with Monte-Carlo techniques have also been used, where an exponential expression for drug release is prescribed \cite{had}. For a comprehensive review of existing mathematical models for mass transfer from polymeric microspheres and transport in tissues, the reader is referred to \cite{ari}. \par 

In a recent work, the problem of a releasing spherical composite capsule has been solved \cite{pon}: therein the external medium
has been confined by a finite length (release distance), say  a cut-off  beyond which the concentration remains sufficiently small and constant. However, in \textit{in-vitro} experiments or in \textit{in-vivo} cases, micro- or nano-spheres are immersed in a bulk ambient medium of size several orders of magnitude larger than that of the sphere itself. Due to the difference of scales, this surrounding medium is considered semi-infinite. \par

In the present paper, a semi-infinite medium is considered around a bi-layered sphere made of an inner core and an outer shell of different drug diffusion coefficients and a rigorous analytic solution based on the Laplace transform is proposed. This approach, used for other biological-oriented models \cite{smg}, in this paper combines ideas presented previously \cite{carr_2016,carr_2017} with some novel features. With the assumption of continuity of diffusive flux between layers, the basic idea of our solution approach is to set the diffusive flux at each of the interlayer surfaces to be equal to an unknown function of time \cite{rod,carr_2017}. This allows the multi-layer problem to be reformulated into a series of coupled single layer problems, which are then solved using the Laplace transform subject to a constraint that enforces that the solutions in each layer satisfy the second specified interlayer condition (after continuity of diffusive flux). The novelty of our approach is that we consider a model consisting of a number of finite layers (spherical shells) together with the semi-infinite outermost layer (ambient medium). The method also avoids the computation of eigenvalues or orthogonal eigenfunction expansions (as used in previous works \cite{pon,carr_2016,carr_2017}), which means that our solution expressions do not require the truncation of infinite series.
 
The remaining sections of this paper are organized in the following way. In the next section, we present the general mathematical model for mass transfer from (into) a multi-layer sphere into (from) a semi-infinite medium. In Section \ref{sec:model}, we consider the special case of a two-layer sphere under the assumption of radial symmetry. This one-dimensional model is then solved using the Laplace transform in Section \ref{sec:solution_procedure} for both the desorbing and absorbing cases. Section \ref{sec:extension} includes extensions of the solution procedure to an arbitrary number of layers 
and to non-uniform initial data. 
 Numerical results and discussion are given in Section \ref{sec:results}.

\section{Mass transfer from/into a composite sphere}
\label{sec:general_model}
\begin{figure}[htb!]
\centering\scalebox{0.4}{\includegraphics{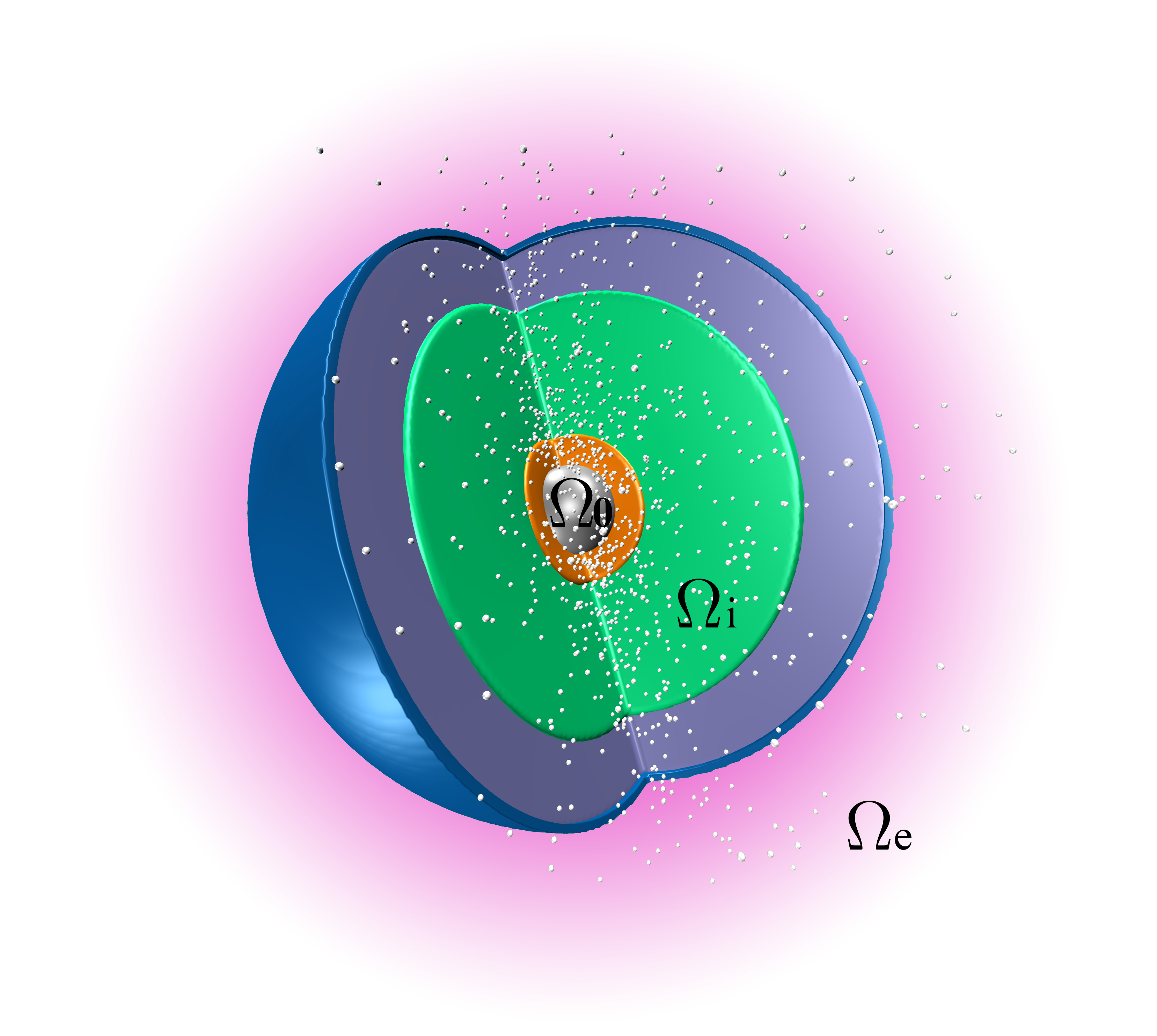}}
\caption{Schematic of a multi-layer microsphere. In the releasing case,
mass is initially loaded in the core $\Omega_0$ and diffuses, through all the intermediate layers $\Omega_i$, into the external medium $\Omega_{e}$ while in the desorbing case the initial mass is present in $\Omega_{e}$ and the direction of diffusion is reversed. Appropriate conditions are imposed at the interfaces between adjacent layers (figure not to scale).}  
\label{fig:microsphere_schematic}
\end{figure}

Consider a multi-layer sphere made of an internal core or depot ($\Omega_0$) enclosed by a number of durable shells ($\Omega_i$, $i=1,2,...,n$, see Fig.~\ref{fig:microsphere_schematic}) constituted of different materials and having specific physico-chemical characteristics. These layers are customized to allow a selective diffusion and better control the transfer rate \cite{tim}. The last shell is immersed in the external ambient medium $\Omega_{e}$ of a large extent (relative to size of the sphere), taken as semi-infinite. In most cases, diffusion is the dominant mechanism of drug transport and, due to the composite nature of the medium, drug kinetics is hard to model and predict. Here we want to study: (i) the mass diffusion from the composite sphere into the
ambient medium (outward flux,  {\em releasing/desorbing sphere}) and (ii) the physically dual process of the absorption from the environment within the sphere (inward flux, {\em absorbing sphere}).

In $\Omega_0$, we assume that the drug dissolution occurs instantaneously compared with that of diffusion \cite{siep}. For diffusion-controlled spheres, the drug release profile is obtained by solving Fick's second law of diffusion: 
\be
{\p c_0 \over \p t} = D_0 \nabla^2 c_0\qquad\text{in $\Omega_0$},  \label{eq1}
\ee
where $c_0$ is the concentration field and $D_0$ is the
diffusion coefficient. Analogously, in the surrounding shells ($\Omega_i$, $i=1,2,...,n$), and in the external medium ($\Omega_{e}$), the following diffusion equations govern the drug transport:
\be
{\p c_i\over \p t} =D_i \nabla^2 c_i\qquad\text{in $\Omega_i$, \qquad $i=1,2,...,n,e$}   \label{eq31} 
\ee
with $c_i$ and $D_i$ denoting the concentration and diffusivity in $\Omega_i$ (possible convection or reaction terms in $\Omega_{e}$ are considered negligible here). The above mechanistic model has been recently introduced in \cite{pon}: in such a study the semi-infinite medium has been truncated at a release distance, an artificial and, in a sense, arbitrary finite cut-off length beyond which all drug is assumed extremely small at a given time and where a perfect sink condition is applied. Similarly to the model presented in \cite{pon}, in this work, we assume:
\begin{enumerate}[(a)]
\item the composite sphere is made of homogeneous enveloping  concentric layers;
\item the diffusivity is constant in each layer;
\item the process is diffusion dominated; 
\end{enumerate}
However, in contrast to \cite{pon} we consider:
\begin{enumerate}[(a)]
\item[(d)] the release medium as semi-infinite and unstirred. 
\end{enumerate}
On the other hand, with respect to the classical single-layer approaches \cite{cra,caj}, we remove some unphysical hypotheses: 
\begin{enumerate}[(i)]
\item $c_0$ is kept constant at  $\p \Omega_0$;
\item $c_e$ is spatially constant in $\Omega_e$ as in a well-stirred medium or in a perfect sink condition.
\end{enumerate}

\bigskip
\noindent\underline{ Modelling interfaces and external coating} \\
At the interfaces between adjacent concentric layers in the capsule, flux continuity holds:
\be
-D_i \nabla c_i  \cdot \bfm{n} =  -D_{i+1} \nabla c_{i+1} \cdot \bfm{n}  \qquad  \text{at $\p \Omega_i \cap \p \Omega_{i+1}$ for $i = 0,\hdots,n-1$,}\label{gh3} 
\ee
with $\bfm{n}$ denoting the surface external unit normal. Additionally, due to partitioning, non-perfect contact exists at the interfaces with the constant ratio of concentration determining the partition coefficient \cite{cra}:
\be
 c_i=\sigma_{i}c_{i+1} \qquad  \text{at $\p \Omega_i \cap \p \Omega_{i+1}$ for $i = 0,\hdots,n-1$,}  \label{gh2}   
\ee
where $\sigma_{i}$ is the drug partition coefficient between layer $i$ and $i+1$.
\bigskip

To prevent fast release, the sphere's outmost shell $\Omega_n$  is protected with a thin coating $\Omega_m$ having diffusivity $D_m$ and a small finite thickness $h$ (Fig. \ref{fig:microsphere_schematic}). This coating shields and preserves the encapsulated drug from degradation and fluid convection, and guarantees a more controlled release \cite{hen}. To model the drug dynamics in $\Omega_m$, we use a simple interface condition between the outermost shell ($\Omega_n$) and the external medium ($\Omega_{e}$) that incorporates the physical properties of the coating as in \cite{pon}:
\be
-D_n \nabla c_n  \cdot \bfm n  = -D_e \nabla c_{e} \cdot \bfm n = P (c_n - \sigma_n c_{e})  \qquad \mbox{at} \,\,
 \p \Omega_n \cap \p \Omega_{e},  \label{eru34}
 \ee
where $P\propto D_m/h$ \, ($\mathrm{m} \mathrm{s}^{-1}$) is the coating mass transfer coefficient and $\sigma_n$ is related to the coating partition coefficients. Note that  Eq.~(\ref{eru34}) includes two limit cases for $P$: when $P=0$ the case of impermeable coating ($\nabla c_n=0$) is obtained, and if $P \rightarrow \infty$ (coating in perfect contact) the form of the other interlayer conditions (\ref{gh2}) is recovered, namely, $c_n= \sigma_n c_e$. 

\section{The core-shell model}
\label{sec:model}
\begin{figure}[!t]
\centering
\includegraphics[scale=0.65,angle=0,trim = 0mm 0mm 0mm 25mm]{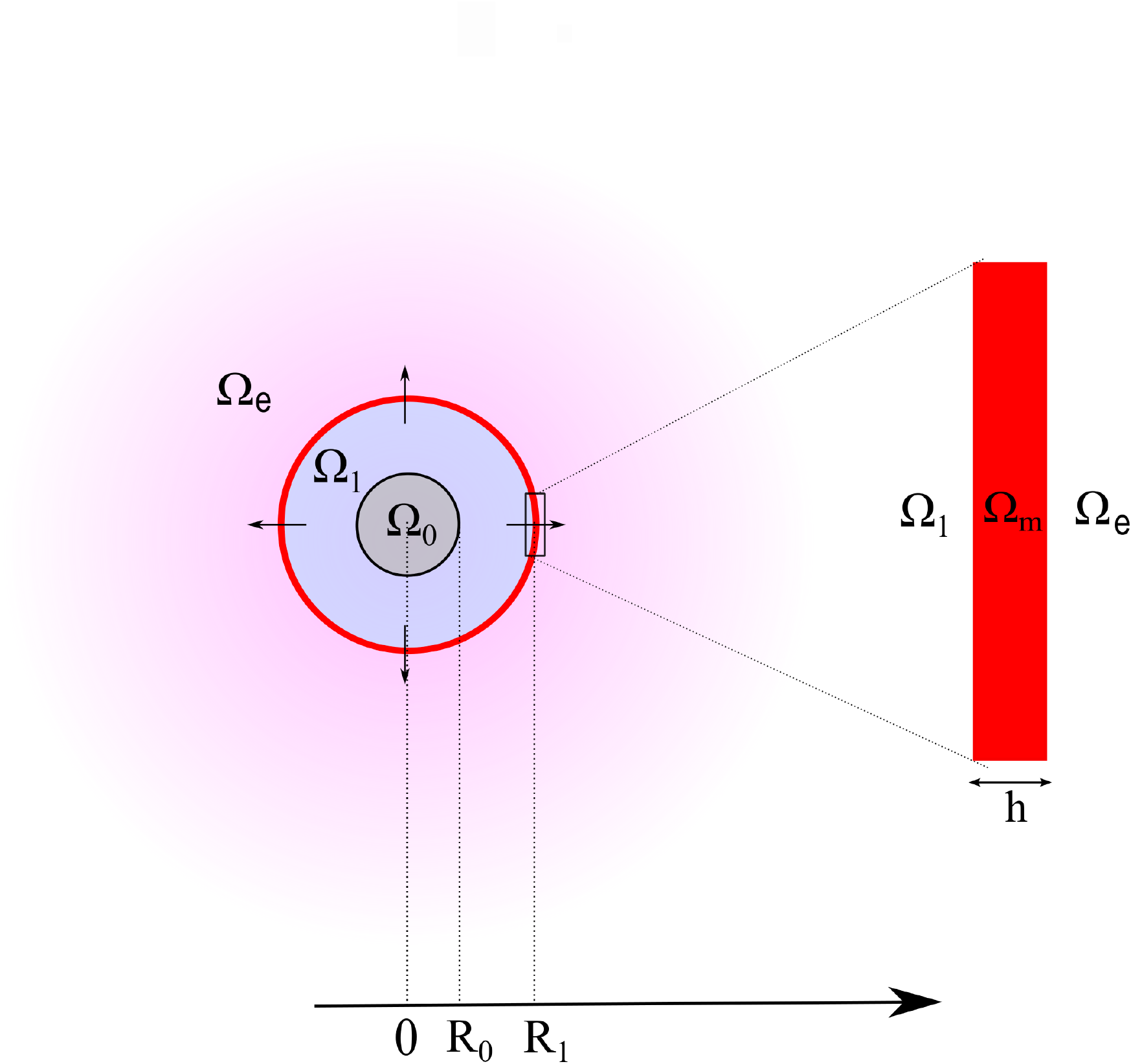}
\caption{Schematic representation of the cross-section 
of the radially symmetric two-layer sphere, comprising an internal core $\Omega_0$, the concentric layer $\Omega_1$ and the thin coating layer $\Omega_m$ (in red, zoomed on the right). This sphere together with the semi-infinite external medium $\Omega_e$, constitutes a three concentric layer system. 
} 
\label{fig:1D_schematic}
\end{figure}

\setcounter{equation}{0}
The two-layer sphere is by far the most representative configuration of absorbing/desorbing system and in this section we consider this special case of $n = 1$ (Fig.~\ref{fig:1D_schematic}): an internal core ($\Omega_0$) encapsulated by a single polymeric shell ($\Omega_1$) surrounded by a ``large'' external medium ($\Omega_{e}$). In other words, the core-shell system is comprised of two concentric spheres of increasing radius immersed in $\Omega_{e}$. Although we consider $n = 1$, it is fairly straightforward to extend the solution methodology presented here to any number of concentric layers as we discuss later in Section \ref{sec:computing_concentration}. We assume the net drug transport occurs in the radial direction only, and therefore we consider a radially-symmetric one-dimensional model (Fig.~\ref{fig:1D_schematic}) as in \cite{pon}. In this case, the general formulation of Section \ref{sec:general_model} is reduced to a three-layer problem, which in 1D radial symmetry reads: 
\begin{alignat}{2}
&{\p c_0 \over \p t} ={D_0 \over r^2} {\p \over \p r}\left(r^2 {\p c_0 \over \p r}\right)&\qquad&\text{in $(0, R_0)$,} \label{erf4}  \\
&{\p c_1 \over \p t} = {D_1 \over r^2} {\p \over \p r}\left(r^2 {\p c_1 \over \p r}\right)&\qquad&\text{in $(R_0, R_1)$} \label{erf5} \\
&{\p c_e \over \p t} = {D_e \over r^2} {\p \over \p r}\left(r^2 {\p c_e \over \p r}\right)&\qquad&\text{in $(R_1, \infty)$,} \label{erf6} \\
&{\p c_0 \over \p r}=0 &\qquad&\text{at $r=0$,} \label{erf7}\\
&-D_0{ \p c_0 \over \p r} = -D_{1}{ \p c_{1} \over \p r} \qquad c_0= \sigma_{0} c_1&\qquad&\text{at $r=R_0$,} \label{erf8}\\
&-D_1{ \p c_1 \over \p r} = -D_{e}{ \p c_{e} \over \p r} = P (c_1 - \sigma_1 c_e) &\qquad&\text{at $r=R_1$,}\label{erf9}\\
&c_e(r,t) = C_e &\qquad& \text{as $r \rightarrow \infty$,} \label{erf0}  
\end{alignat}
where $r$ is the radial coordinate. 

The initial conditions are:
\begin{alignat}{2}
\label{eq:ic0}
&c_0(r,0)=C_0, &\qquad& \text{in $(0,R_0)$,}\\
\label{eq:ic1}
&c_1(r,0)=C_1, &\qquad& \text{in $(R_0,R_1)$,}\\
\label{eq:ic2}
&c_e(r,0)=C_e, &\qquad& \text{in $(R_1,\infty)$,}
\end{alignat} 
where $C_0$, $C_1$ and $C_e$ are specified constants. For a desorbing/releasing sphere we take $C_{0}>0$ and $C_1 = C_e =0$ while for an absorbing sphere we take $C_0 = C_1 = 0$ and $C_{e}>0$.

\section{Solution procedure}
\label{sec:solution_procedure}
\setcounter{equation}{0}

We now present our solution approach for solving the core-shell model (\ref{erf4})--(\ref{eq:ic2}). As a first step, we normalize the variables, the parameters and the equations via the change of variables:
\be
 r \rightarrow {r \over R_1 }, \qquad  t \rightarrow { D_{\max} \over R_1^2 t }, \qquad  c_i  \rightarrow {c_i \over C_{\mathrm{max}}},  \qquad  C_i  \rightarrow {C_i \over C_{\mathrm{max}}},  \qquad  \text{for $i=0,1,e$,}
\ee
and by redefining the nondimensional constants:
\be
R_{i} \rightarrow  { R_{i}  \over R_{1}},\qquad D_i \rightarrow {D_i \over  D_{\max}},\qquad   P \rightarrow {P R_{1} \over D_{\max}},     \label{gh6}
\ee
where $C_{\max} := \max ({C_0, C_1, C_e})$ and $D_{\max} := \max({D_0, D_1, D_e})$.

\subsection{Reformulating the problem}
\label{sec:reformulating}
Start by defining the unknown mass fluxes, $g_{0}(t)$ and $g_{1}(t)$ \cite{rod,carr_2017}:
\begin{alignat}{2}
\label{eq:g0_R0}
&g_{0}(t) :=  -D_{0}\frac{\partial c_{0}}{\partial r} = -D_{1}\frac{\partial c_{1}}{\partial r}&\qquad&\text{at $r = R_{0}$,}\\
\label{eq:g1_R1}
&g_{1}(t) :=  -D_{1}\frac{\partial c_{1}}{\partial r} = -D_{e}\frac{\partial c_{e}}{\partial r}&\qquad&\text{at $r = R_{1}$.}
\end{alignat}
Neglect the partition interface conditions for now, and consider the decoupled problems arising from (\ref{erf4})--(\ref{erf0}) \cite{rod,carr_2017}:
\begin{enumerate}[(i)]
\item Internal core ($\Omega_{0}$):
\begin{equation}
\label{eq:c0_problem}
\begin{alignedat}{2}
&{\p c_0 \over \p t} = {D_0 \over r^2} {\p \over \p r}\left(r^2 {\p c_0 \over \p r}\right)&\qquad&\text{in $(0, R_0)$,}\\
&c_0(r,0)=C_{0}&\qquad&\text{at $t = 0$,}\\
&\frac{\partial c_{0}}{\partial r} = 0&\qquad&\text{at $r = 0$,} \\
&-D_{0}\frac{\partial c_{0}}{\partial r} = g_{0}(t) &\qquad&\text{at $r= R_{0}$.}
\end{alignedat}
\end{equation}
\item Polymeric shell ($\Omega_{1}$):
\begin{equation}
\label{eq:c1_problem}
\begin{alignedat}{2}
&{\p c_1 \over \p t} = {D_1 \over r^2} {\p \over \p r}\left(r^2 {\p c_1 \over \p r}\right)&\qquad&\text{in $(R_0, R_1)$,}\\
&c_1(r,0)=C_{1}&&\text{at $t = 0$,}\\
&-D_{1}\frac{\partial c_{1}}{\partial r} = g_{0}(t)&&\text{at $r = R_{0}$,}\\
&-D_{1}\frac{\partial c_{1}}{\partial r} = g_{1}(t)&&\text{at $r = R_{1}$.}
\end{alignedat}
\end{equation}
\item External medium ($\Omega_{e}$):
\begin{equation}
\label{eq:c2_problem}
\begin{alignedat}{2}
&{\p c_e \over \p t} = {D_e \over r^2} {\p \over \p r}\left(r^2 {\p c_e \over \p r}\right)&\qquad&\text{in $(R_1,\infty)$,}\\
&c_e(r,0)=C_{e} &&\text{at $t = 0$,}\\
&-D_{e}\frac{\partial c_{e}}{\partial r} = g_{1}(t)&&\text{at $r = R_{1}$,}\\
&c_e(r,t)=C_{e}&&\text{at $r \rightarrow \infty$.} 
\end{alignedat}
\end{equation}
\end{enumerate}

\subsection{Computing the concentration}
\label{sec:computing_concentration}
Let us consider first the solution in the internal core ($\Omega_{0}$). The Laplace transform of (\ref{eq:c0_problem}) yields the following boundary value problem for $\overline{c}_{0}(r,s) := \mathcal{L}\{c_{0}(r,t)\}$:
\begin{alignat}{2}
\label{bvp_eq}
&s\overline{c}_0 - C_{0} = {D_0 \over r^2} {d\over dr}\left(r^2 {d \overline{c}_0 \over dr}\right)&\qquad&\text{in $(0, R_0)$,}\\
\label{bvp_lbc}
&\frac{d\overline{c}_{0}}{dr} = 0&&\text{at $r = 0$,}\\
\label{bvp_rbc}
&-D_{0}\frac{d\overline{c}_{0}}{dr} = \overline{g}_{0}(s)&&\text{at $r = R_{0}$.}
\end{alignat}
The general solution of (\ref{bvp_eq}) is
\begin{align}
&\overline{c}_{0}(r,s) = \frac{C_{0}}{s} + A\frac{\sinh\left(\mu_{0}(s) r\right)}{r} + B\frac{\cosh\left(\mu_{0}(s) r\right)}{r},
\end{align}
where $\mu_{0}(s) := \sqrt{s/D_{0}}$. Now, consider:
\begin{multline}
\label{eq:dcbar0}
\frac{d\overline{c}_{0}}{dr}(r,s) = A\left[\frac{\cosh\left(\mu_{0}(s) r\right)\mu_{0}(s)}{r} - \frac{\sinh\left(\mu_{0}(s) r\right)}{r^2}\right]\\ + B\left[\frac{\cosh\left(\mu_{0}(s) r\right)\mu_{0}(s)}{r} - \frac{\cosh\left(\mu_{0}(s) r\right)}{r^2}\right].
\end{multline}
The boundary condition (\ref{bvp_lbc}) requires $B = 0$. Using (\ref{eq:dcbar0}) with $B = 0$ and applying (\ref{bvp_rbc}) allows $A$ to be identified:
\begin{align}
A = - \frac{R_{0}^{2}\overline{g}_{0}(s)}{D_{0}\left[\cosh\left(\mu_{0}(s)R_{0}\right)\mu_{0}(s)R_{0} - \sinh\left(\mu_{0}(s)R_{0}\right)\right]}.
\end{align}
In summary, the solution of (\ref{bvp_eq})--(\ref{bvp_rbc}) is given by:
\begin{align}
\label{eq:c0bar}
\overline{c}_{0}(r,s) = \frac{C_{0}}{s} + a_{0,1}(r,s)\overline{g}_{0}(s),
\end{align}
where
\begin{align}
\label{eq:a00}
a_{0,1}(r,s) :=  - \frac{R_{0}^{2}\sinh\left(\mu_{0}(s) r\right)}{rD_{0}\left[\cosh\left(\mu_{0}(s)R_{0}\right)\mu_{0}(s)R_{0} - \sinh\left(\mu_{0}(s)R_{0}\right)\right]}.
 \end{align}
Carrying out a similar process for the polymeric shell ($\Omega_{1}$) and external medium ($\Omega_{e}$), we obtain the following expressions for the Laplace transforms of $c_{1}(r,t)$ and $c_{e}(r,t)$:
\begin{align}
\label{eq:c1bar}
&\overline{c}_{1}(r,s) = \frac{C_{1}}{s} + a_{1,0}(r,s)\overline{g}_{0}(s) + a_{1,1}(r,s)\overline{g}_{1}(s),\\ 
\label{eq:c2bar}
&\overline{c}_{e}(r,s) = \frac{C_{e}}{s} + a_{e,0}(r,s)\overline{g}_{1}(s),
\end{align}
where:
\begin{align}
\label{eq:a10}
&a_{1,0}(r,s) := \frac{R_{0}^2\left[\mu_{1}(s)R_{1}\cosh\left(\mu_{1}(s)\left(r-R_{1}\right)\right) + \sinh\left(\mu_{1}(s)\left(r-R_{1}\right)\right)\right]}{r\left[D_{1}\mu_{1}(s)\Delta R_{1} \cosh\left(\mu_{1}(s)\Delta R_{1}\right)+ (sR_{1}R_{0}-D_{1})\sinh\left(\mu_{1}(s)\Delta R_{1}\right)\right]},\\
\label{eq:a11}
&a_{1,1}(r,s) := \frac{-R_{1}^{2}\left[\mu_{1}(s)R_{0}\cosh\left(\mu_{1}(s)\left(r-R_{0}\right)\right) + \sinh\left(\mu_{1}(s)\left(r-R_{0}\right)\right)\right]}{r\left[D_{1}\mu_{1}(s)\Delta R_{1} \cosh\left(\mu_{1}(s)\Delta R_{1}\right)+ (sR_{1}R_{0}-D_{1})\sinh\left(\mu_{1}(s)\Delta R_{1}\right)\right]},\\
\label{eq:a21}
&a_{e,0}(r,s) := \frac{R_{1}^{2}\exp\left(-\mu_{e}(s)(r-R_{1})\right)}{rD_{e}\left[1 + \mu_{e}(s)R_{1}\right]}. 
\end{align}
In Eqs (\ref{eq:a10})--(\ref{eq:a21}) we have set $\Delta R_{1} := R_{1}-R_{0}$, $\mu_{1}(s) := \sqrt{s/D_{1}}$ and $\mu_{e}(s) := \sqrt{s/D_{e}}$ for ease of notation.

Applying the inverse Laplace transform to Eqs (\ref{eq:c0bar}), (\ref{eq:c1bar}) and (\ref{eq:c2bar}) yields the concentration in each layer:
\begin{align}
\label{eq:c0soln}
c_{0}(r,t) &= \mathcal{L}^{-1}\left\{\bar c_0 (r,s)\right\} = C_{0} + \mathcal{L}^{-1}\left\{a_{0,1}(r,s)\overline{g}_{0}(s)\right\},\\
\label{eq:c1soln}
c_{1}(r,t) &= \mathcal{L}^{-1}\left\{\bar c_1(r,s)\right\} = C_{1} + \mathcal{L}^{-1}\left\{a_{1,0}(r,s)\overline{g}_{0}(s)\right\} + \mathcal{L}^{-1}\left\{a_{1,1}(r,s)\overline{g}_{1}(s)\right\},\\
\label{eq:c2soln} 
c_e(r,t) &= \mathcal{L}^{-1}\left\{\bar c_e(r,s)\right\} = C_{e} + \mathcal{L}^{-1}\left\{a_{e,0}(r,s)\overline{g}_{1}(s)\right\}. 
\end{align}
To evaluate the solutions (\ref{eq:c0soln})--(\ref{eq:c2soln}) at a given time $t$, the following quadrature formula \cite{trefethen_2006} is used to calculate the inverse Laplace transforms:
\begin{align}
\label{numlap}
\mathcal{L}^{-1}\left\{a_{ij}(r,s)\overline{g}_{j}(s)\right\} \approx -2\mathfrak{Re}\left\{\sum_{k=1}^{N/2}f_{2k-1}\frac{a_{ij}(r,s_{k})\overline{g}_{j}(s_{k})}{t}\right\},
\end{align}
for $i = 0,1,e$ and $j = 0,1$, where $s_{k}:=z_{2k-1}/t$ and $\mathfrak{Re}\{\cdot\}$ denotes the real part. The constants $f_{2k-1}$ and $z_{2k-1}$, $k = 1,\hdots,N/2$, are defined as the residues and poles, respectively, of the best $(N,N)$ rational approximation to the exponential function on the negative real line, computed via the Carath\'eodroy-Fej\'er method \cite{trefethen_2006}. All results generated in this paper are computed by setting $N = 14$ \cite{carr_2016}. The computation (\ref{numlap}) requires evaluation of the unknown functions $\overline{g}_{0}(s)$ and $\overline{g}_{1}(s)$ at $s_{k} = z_{2k-1}/t$. To this aim, we solve the following linear system derived by taking Laplace transforms of the (as yet unused) interface conditions in Eqs (\ref{erf8}) and (\ref{erf9}):
\begin{align}
&\overline{c}_{0}(R_0,s_{k}) - \sigma_{0}\overline{c}_{1}(R_0,s_k) = 0,  \label{ce1} \\
&\overline{c}_{1}(R_1,s_{k}) - \sigma_1 \overline{c}_{e}(R_{1},s_k) = \frac{1}{P}\overline{g}_1(s_k)  \label{ce2}.
\end{align}
Substitution of (\ref{eq:c0bar}), (\ref{eq:c1bar}) and (\ref{eq:c2bar}) into (\ref{ce1}) and (\ref{ce2}) produces a linear system of equations:
\begin{gather}
\label{matsy}
\mathbf{A}\mathbf{x} = \mathbf{b},
\end{gather}
where:
\begin{gather}
\label{eq:matsy_A}  
\mathbf{A} = \left(\begin{matrix} a_{0,1}(R_{0},s_{k})-\sigma_{0}a_{1,0}(R_0,s_{k}) & -\sigma_{0}a_{1,1}(R_0,s_{k})\\ a_{1,0}(R_1,s_{k})& a_{1,1}(R_1,s_{k})- \sigma_1 a_{e,0}(R_1,s_{k}) - P^{-1} \end{matrix}\right),\\
\mathbf{x} = \left(\begin{matrix}\overline{g}_{0}(s_{k})\\ \overline{g}_{1}(s_{k})\end{matrix}\right),\\
\label{eq:matsy_b}
\mathbf{b} = \left(\begin{matrix}\left(\sigma_{0}C_{1}-C_{0}\right)/s_{k}\\ \left(\sigma_1 C_{e} - C_{1}\right)/s_{k}\end{matrix}\right).
\end{gather}
Solving the linear system (\ref{matsy}) for $\mathbf{x}$ yields the evaluations $\overline{g}_{0}(s_{k})$ and $\overline{g}_{1}(s_{k})$, appearing in the numerical inverse Laplace transforms (\ref{numlap}), which completes the solution.

\subsection{Computing the mass}
The drug mass in all concentric layers as a function of time is defined as the volume integral of the concentration, which simplifies to the following formulas under the assumption of radial symmetry \cite{pon}:
\begin{gather}
\label{eq:mass1}
M_0(t)= 4\pi\int_{0}^{R_0} r^{2}c_0(r,t)\,\mathrm{d}r, \qquad 
M_1(t) =  4\pi\int_{R_0}^{R_1}  r^{2}c_1(r,t)\,\mathrm{d}r,\\
\label{eq:mass2}
M_e(t) = 4\pi\int_{R_1}^{\infty}  r^{2}c_e(r,t)\,\mathrm{d}r.
\end{gather}
To evaluate the integrals, the solution expressions (\ref{eq:c0soln})--(\ref{eq:c2soln}) are substituted into (\ref{eq:mass1})--(\ref{eq:mass2}) and the resulting integrals computed numerically. Using the initial conditions (\ref{eq:ic0})--(\ref{eq:ic2}), we have: 
\be
M_0(0)=\frac{4}{3}\pi R_{0}^{3}C_{0}, \qquad M_1(0)=\frac{4}{3}\pi(R_1^3-R_0^3)C_1, \qquad M_e(0)=\begin{cases} 0 & \text{if $C_{e} = 0$}\\ \infty & \text{if $C_{e}= 1$}. \end{cases} \label{eq5}
\ee
For the desorbing case with initial data $C_{0}=1$ and $C_{1} = C_{e} = 0$:
\begin{align}
&\lim\limits_{t \rightarrow \infty} M_0(t) = \lim\limits_{t \rightarrow \infty} M_1(t) = 0,\qquad \lim\limits_{t \rightarrow \infty} M_e(t)= \frac{4}{3}\pi R_{0}^{3} C_{0}=M_0(0),  \label{eq6}
\end{align}
while for the absorbing case with initial conditions $C_{0} = C_{1} = 0$ and $C_{e}=1$:
\begin{align}
&\lim\limits_{t \rightarrow \infty} M_0(t) = \frac{4}{3}\pi R_{0}^{3}C_{e},\qquad
\lim\limits_{t \rightarrow \infty} M_1(t) = \frac{4}{3}\pi (R_{1}^{3}-R_{0}^{3})C_{e}.  \label{mass_abs}
\end{align}

Note that in the desorbing case, as described in Eqn (\ref{eq6}), all the initial (finite) mass is transferred outside to the external medium ($\Omega_{e}$), while in the absorbing sphere an initial (infinite) mass is initially given in $\Omega_e$ and residual mass remains there for all time.


\section{Extension of the solution procedure}
\setcounter{equation}{0}
\label{sec:extension}
\underline{Arbitrary number of concentric layers}\\
We now discuss extension of the analytical solution derived in Section \ref{sec:computing_concentration} to the case of a spherical core enveloped by an arbitrary number $n$ of concentric spheres with increasing radius, such that $R_0 < R_1 < \hdots < R_n$. In this case, the analogue of the model (\ref{erf4})--(\ref{eq:ic2}) is given by:
\begin{alignat*}{2}
&{\p c_0 \over \p t} = {D_0 \over r^2} {\p \over \p r}\left(r^2 {\p c_0 \over \p r}\right)&\qquad\quad&\text{in $(0, R_0)$,}  \\
&{\p c_i \over \p t} = {D_i \over r^2} {\p \over \p r}\left(r^2 {\p c_i \over \p r}\right)&&\text{in $(R_{i-1}, R_i)$ for $i = 1,\hdots,n$,}\\
&{\p c_{e} \over \p t} = {D_e \over r^2} {\p \over \p r}\left(r^2 {\p c_{e} \over \p r}\right)&&\text{in $(R_n, \infty)$,}
\end{alignat*}
subject to the boundary, interface and initial conditions:
\begin{alignat*}{2}
&{\p c_0 \over \p r}=0 &\qquad\quad&\text{at $r=0$,}\\
&-D_{i}{ \p c_i \over \p r} = -D_{i+1}{ \p c_{i+1} \over \p r}, \qquad c_i= \sigma_{i}c_{i+1}&&\text{at $r=R_i$ for $i = 0,\hdots,n-1$,}\\
&-D_{n}{ \p c_n \over \p r} = -D_{e}{ \p c_{e} \over \p r} = P(c_{n} - \sigma_n c_{e}) &&\text{at $r=R_n$,}\\
&c_{e}(r,t) = C_{e} && \text{as $r \rightarrow \infty$,}\\
&c_0(r,0) = C_0 && \text{in $(0, R_0)$,}\\
&c_i(r,0) = C_i && \text{in $(R_{i-1}, R_i)$ for $i = 1,\hdots,n$,}\\ 
&c_e(r,0) = C_e && \text{in $(R_n, \infty)$.}  
\end{alignat*}
The general problem above can be solved using an identical procedure to that described for the case study in Section \ref{sec:model}. Firstly, the problem is reformulated into a series of $n+2$ single layer problems in a similar manner to that described in Section \ref{sec:reformulating} by first defining the following $n+1$ unknown mass fluxes:
\begin{alignat}{2}
\label{eq:g0_R0}
&g_{i}(t) :=  -D_{i}\frac{\partial c_{i}}{\partial r} = -D_{i+1}\frac{\partial c_{i+1}}{\partial r}&\qquad&\text{at $r = R_{i}$ for $i = 0,\hdots,n-1$,}\\
&g_{n}(t) :=  -D_{n}\frac{\partial c_{n}}{\partial r} = -D_{e}\frac{\partial c_{e}}{\partial r}&\qquad&\text{at $r = R_{n}$.}
\end{alignat}
Applying now the same Laplace transform methodology described in Section \ref{sec:computing_concentration}, yields the solutions:
\begin{align}
c_{0}(r,t) &= C_{0} + \mathcal{L}^{-1}\left\{a_{0,1}(r,s)\overline{g}_{0}(s)\right\},\\
c_{i}(r,t) &= C_{i} + \mathcal{L}^{-1}\left\{a_{i,0}(r,s)\overline{g}_{i-1}(s)\right\} + \mathcal{L}^{-1}\left\{a_{i,1}(r,s)\overline{g}_{i}(s)\right\}\qquad\text{for $i = 1,\hdots,n$,}\\
c_{e}(r,t) &= C_{e} + \mathcal{L}^{-1}\left\{a_{e,0}(r,s)\overline{g}_{n}(s)\right\}. 
\end{align}
The form of $a_{0,1}(r,s)$ remains unchanged from (\ref{eq:a00}) while the indexes in (\ref{eq:a10})--(\ref{eq:a21}) are modified to account for the arbitrary number of layers:
\begin{align*}
&a_{i,0}(r,s) := \frac{R_{i-1}^2\left[\mu_{i}(s)R_{i}\cosh\left(\mu_{i}(s)\left(r-R_{i}\right)\right) + \sinh\left(\mu_{i}(s)\left(r-R_{i}\right)\right)\right]}{r\left[D_{i}\mu_{i}(s)\Delta R_{i} \cosh\left(\mu_{i}(s)\Delta R_{i}\right)+ (sR_{i}R_{i-1}-D_{i})\sinh\left(\mu_{i}(s)\Delta R_{i}\right)\right]},\\
&a_{i,1}(r,s) := -\frac{R_{i}^{2}\left[\mu_{i}(s)R_{i-1}\cosh\left(\mu_{i}(s)\left(r-R_{i-1}\right)\right) + \sinh\left(\mu_{i}(s)\left(r-R_{i-1}\right)\right)\right]}{r\left[D_{i}\mu_{i}(s)\Delta R_{i} \cosh\left(\mu_{i}(s)\Delta R_{i}\right)+ (sR_{i}R_{i-1}-D_{i})\sinh\left(\mu_{i}(s)\Delta R_{i}\right)\right]},\\
&a_{e,0}(r,s) := \frac{R_{n}^{2}\exp\left(-\mu_{e}(s)(r-R_{n})\right)}{rD_{e}\left[1 + \mu_{e}(s)R_{n}\right]},
\end{align*}
with $\Delta R_{i} := R_{i}-R_{i-1}$ and $\mu_{i}(s) := \sqrt{s/D_{i}}$. 

The inverse Laplace transforms are computed using (\ref{numlap}) with the evaluations $\overline{g}_{j}(s_k)$ ($j = 0,1,\hdots,n$) satisfying an extended version of the system (\ref{matsy}) with dimension $(n+1)\times (n+1)$, where $\mathbf{A}$ is a tridiagonal matrix. The entries of $\mathbf{A}$, $\mathbf{x}$ and $\mathbf{b}$ are defined as follows:
\begin{alignat*}{2}
&A_{j,j}= \widetilde{a}_{j-1}(R_{j},s_{k}) &\qquad& \text{for $j=1,...,n+1$},\\
&A_{j,j-1}={a}_{j-1,0}(R_{j},s_{k}) &\qquad& \text{for $j=2,...,n+1$,}\\
&A_{j,j+1}=-\sigma_{j-1}{a}_{j+1,1}(R_{j},s_{k}) &\qquad& \text{for $j=1,...,n$,}\\
&x_{j}= \overline{g}_{j-1}(s_{k}) &\qquad& \text{for $j=1,...,n+1$,} \\
&b_{j}= \left(\sigma_{j-1}C_{j}-C_{j-1}\right)/s_{k} &\qquad& \text{for $j=1,...,n$},\\
&b_{n+1} = \left(\sigma_n C_{e}-C_{n}\right)/s_{k},
\end{alignat*}
where $\widetilde{a}_{i}(R_{i},s_{k}) := a_{i,1}(R_{i},s_{k})-\sigma_{i}a_{i+1,0}(R_{i},s_{k})$ for $i = 0,1,\hdots,n-1$ and $\widetilde{a}_{n}(R_{n},s_{k}) := a_{n,1}(R_{n},s_{k})-\sigma_n a_{e,0}(R_n,s_{k}) - P^{-1}$.\\

\noindent\underline{Non-uniform initial data}\\
So far, in both the mathematical modelling and solution procedure, we have assumed uniform initial conditions in each concentric layer, i.e., $C_{0}$, $C_{1}$ and $C_{2}$ in Eqs (\ref{eq:ic0})--(\ref{eq:ic2}) are constants. In this section, we revisit the two-layer sphere and expand the solution procedure outlined in Section \ref{sec:solution_procedure} to spatially-dependent initial conditions. To explain the process, consider the model described in Section \ref{sec:model} with Eq.~(\ref{eq:ic0}) replaced with the non-uniform initial condition: 
\begin{align}
c_{0}(r,0) = C_{0}(r)\qquad \mbox{in}\, \, (0, R_0).
\end{align}
In this case, Eq.~(\ref{bvp_eq}) becomes: 
\begin{alignat}{2}
\label{eq:c0eq_nonuniform}
&s\overline{c}_0 - C_{0}(r) = {D_0 \over r^2} {d\over dr}\left(r^2 {d \overline{c}_0 \over dr}\right)&\qquad&\text{in $(0, R_0)$}.
\end{alignat}
The general solution of (\ref{eq:c0eq_nonuniform}) is now:
\begin{align}
&\overline{c}_{0}(r,s) = \overline{c}_{0}^{(p)}(r,s) + A\frac{\sinh\left(\mu_{0}(s) r\right)}{r} + B\frac{\cosh\left(\mu_{0}(s) r\right)}{r},
\end{align}
where the particular solution, $\overline{c}_{0}^{(p)}(r,s)$, can be derived using the method of variation of parameters \cite{tri}:
\begin{align}
\overline{c}_{0}^{(p)}(r,s) = \frac{1}{rD_{0}\mu_{0}(s)}\int_{0}^{r}uC_{0}(u)\sinh\left(\mu_{0}(s)(u-r)\right)\,\mathrm{d}u.
\end{align}
Applying the boundary conditions (\ref{bvp_lbc})--(\ref{bvp_rbc}) produces the modified form of (\ref{eq:c0bar}) for the case of non-uniform initial data in the first layer:
\begin{align}
\overline{c}_{0}(r,s) = \widetilde{c}_{0}(r,s) + a_{0,1}(r,s)\overline{g}_{0}(s),
\end{align}
where $a_{0,0}(r,s)$ is as defined in Eq.~(\ref{eq:a00}) and
\begin{multline}
\widetilde{c}_{0}(r,s) = \overline{c}_{0}^{(p)}(r,s) - \frac{a_{0,1}(r,s)}{R_{0}D_{0}\mu_{0}(s)}\left[\int_{0}^{R_{0}}uC_{0}(u)\cosh\left(\mu_{0}(s)(u-R_{0})\right)\mu_{0}(s)\,\mathrm{d}u\right.\\\left. + \frac{1}{R_{0}}\int_{0}^{R_{0}}uC_{0}(u)\sinh\left(\mu_{0}(s)(u-R_{0})\right)\,\mathrm{d}u\right].
\end{multline}
The inverse Laplace transform yields the modified solution in the first layer:
\begin{align}
c_{0}(r,t) = \mathcal{L}^{-1}\left\{\overline{c}_{0}(r,s)\right\} = \mathcal{L}^{-1}\left\{\widetilde{c}_{0}(r,s)\right\} + \mathcal{L}^{-1}\left\{a_{0,1}(r,s)\overline{g}_{0}(s)\right\},
\end{align}
where $\mathcal{L}^{-1}\left\{\widetilde{c}_{0}(r,s)\right\}$ can be evaluated using the quadrature formula (\ref{numlap}) as follows:
\begin{align}
\mathcal{L}^{-1}\left\{\widetilde{c}_{0}(r,s)\right\} \approx -2\mathfrak{Re}\left\{\sum_{k=1}^{N/2}f_{2k-1}\frac{\widetilde{c}_{0}(r,s_{k})}{t}\right\}.
\end{align}
The evaluations $\overline{g}_{0}(s_{k})$ and $\overline{g}_{1}(s_{k})$ are computed as in Section \ref{sec:computing_concentration} with the exception that (\ref{eq:matsy_b}) is replaced with:
\begin{align}
\mathbf{b} = \left(\begin{matrix}\sigma_{0}C_{1}/s_{k}-\widetilde{c}_{0}(R_{0},s_{k})\\ \left(\sigma_n C_{e} - C_{1}\right)/s_{k}\end{matrix}\right).
\end{align}

\section{Results and discussion}
\label{sec:results}
\setcounter{equation}{0} 

\subsection{Solution verification for a homogeneous sphere}  
\label{sec:homogeneous_sphere}
We first assess our Laplace transform solution (\ref{eq:c0soln})--(\ref{eq:c2soln}) using the homogeneous analogue of the core shell-model (\ref{erf4})--(\ref{eq:ic2}):
\begin{alignat*}{2}
&\frac{\partial c}{\partial t} = \frac{D}{r^{2}}\frac{\partial}{\partial r}\left(r^{2}\frac{\partial c}{\partial r}\right)&\qquad&\text{in $(0, \infty)$,}\\
&c(r,0)=f(r) &&\text{in $(0, \infty)$,}\\
&{\p c \over \p r}=0 &&\text{at $r=0$,}\\
&c(r,t) = C_{e} && \text{as $r \rightarrow \infty$,} 
\end{alignat*}
where we have a single-sphere of radius $R$ and take 
\begin{gather*}
f(r) = \begin{cases} 
C_0 & \text{for $r < R$}\\ 
C_e & \text{for $r > R$,}
\end{cases}
\end{gather*}
with $C_0=1$ and $C_e=0$ for the desorbing case and $C_0=0$ and $C_e=1$ for the absorbing case.

\begin{figure}[!b]
\centering
\subfloat[Desorbing]{\includegraphics[width=0.5\textwidth]{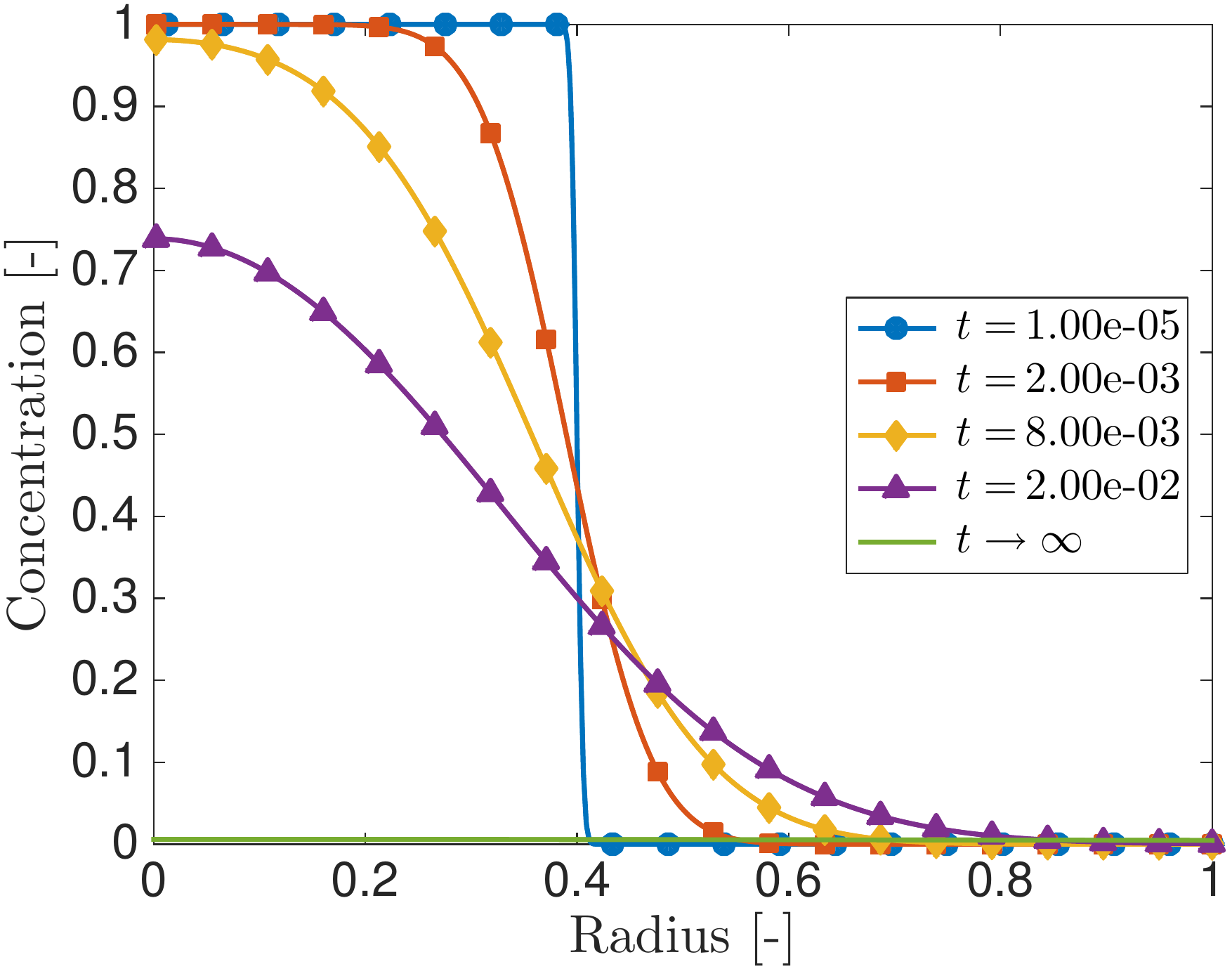}}
\subfloat[Absorbing]{\includegraphics[width=0.5\textwidth]{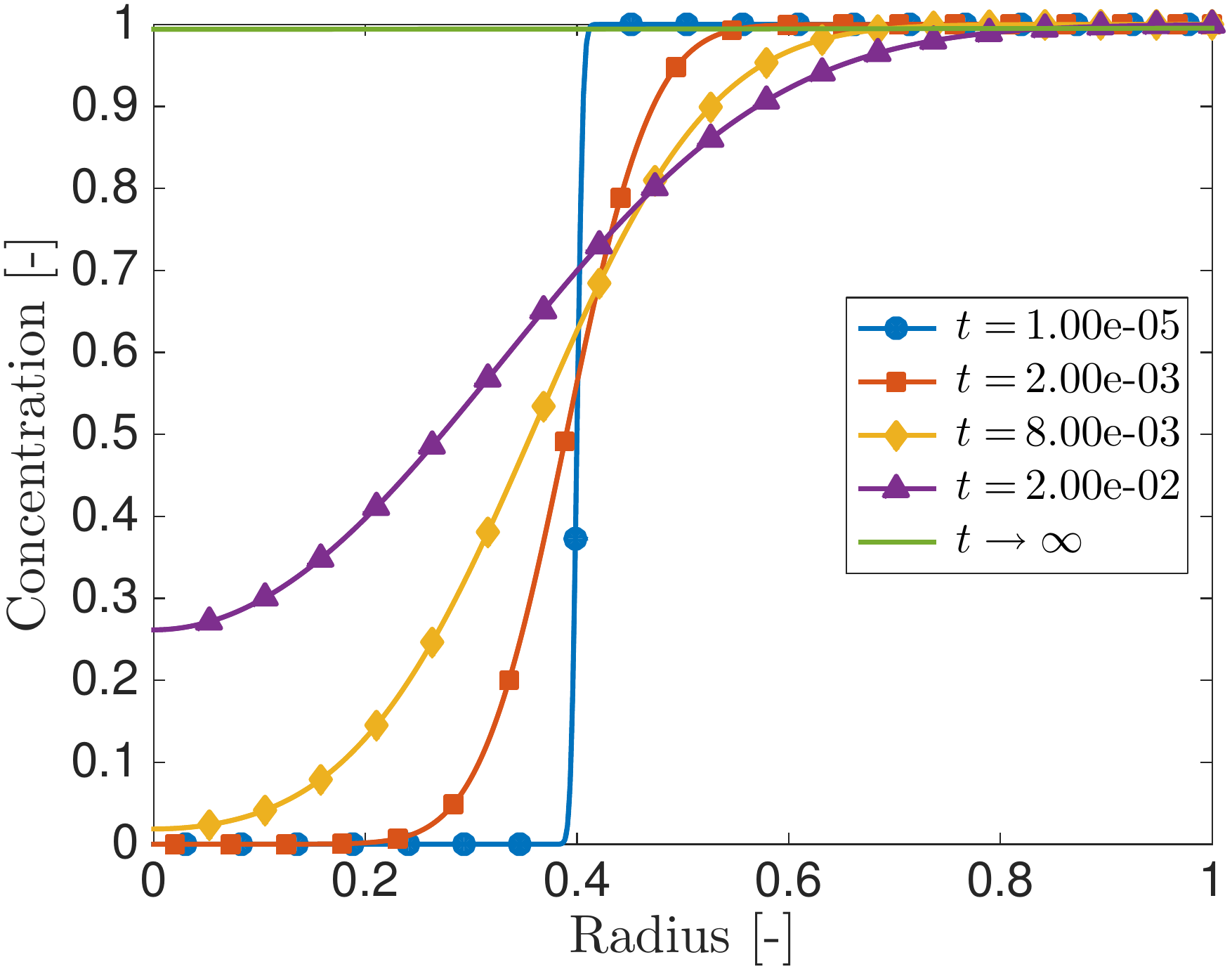}}
\caption{Concentration profiles for the homogeneous sphere test problem 
at several (dimensionless) times for (a) desorbing and (b) absorbing cases. The continuous lines denote the concentration profiles obtained using the Laplace transform solution (\ref{eq:c0soln})--(\ref{eq:c2soln}) while the markers represent the classical solutions (\ref{eq:classical_soln_desorbing})--(\ref{eq:classical_soln_absorbing}).}
\label{fig:homogeneous_sphere}
\end{figure}

The exact solution of this simplified problem is well-known for an arbitrary initial function $f(r)$ \cite{pol}. Substituting the specific forms of $f(r)$ into this exact solution yields for the desorbing case:
\begin{align}
\label{eq:classical_soln_desorbing}
c(r,t) = \frac{1}{2r\sqrt{\pi D t}}\int_{0}^R\xi \left\{\exp\left[-\frac{(r-\xi)^{2}}{4Dt}\right] - \exp\left[-\frac{(r+\xi)^{2}}{4Dt}\right]\right\}\,\mathrm{d}\xi,
\end{align}
and for the absorbing case:
\begin{align}
\label{eq:classical_soln_absorbing}
c(r,t) = \frac{1}{2r\sqrt{\pi D t}}\int_R^{\infty}\xi \left\{\exp\left[-\frac{(r-\xi)^{2}}{4Dt}\right] - \exp\left[-\frac{(r+\xi)^{2}}{4Dt}\right]\right\}\,\mathrm{d}\xi,
\end{align} 
where the solutions are valid for $r >0$ in both cases.

In Fig.~\ref{fig:homogeneous_sphere}, we compare the above classical solutions to our Laplace transform solution (\ref{eq:c0soln})--(\ref{eq:c2soln}) for dimensionless values of $D = 1$ and $R = 0.4$. To solve the homogeneous model, we use the following choices of parameters in the core shell-model (\ref{erf4})--(\ref{eq:ic2}): $D_{0}=D_{1}=D_{e} = D$, $R_{1} = R$ and $P\rightarrow\infty$ (i.e. $P^{-1} = 0$ in Eq. (\ref{eq:matsy_A}),  with $C_{0} = C_{1} = 1$ and $C_{e} = 0$ for the desorbing case and $C_{1}=C_{2}=0$ and $C_{e} = 1$ for the absorbing case. Note that the concentration profiles are depicted on a truncated finite domain, $0\leq r\leq 1$, as beyond $r = 1$ the solution is effectively constant. Clearly, at all times shown, both the Laplace transform solution (\ref{eq:c0soln})--(\ref{eq:c2soln}) (continuous lines in Fig. \ref{fig:homogeneous_sphere}) and classical solutions (\ref{eq:classical_soln_desorbing})--(\ref{eq:classical_soln_absorbing}) (markers in Fig. \ref{fig:homogeneous_sphere}) are in excellent agreement. In summary, these results confirm numerically that our analytical solution correctly reduces to the exact solution of the homogeneous model when $D_{0} = D_{1} = D_{e} = D$,\enspace$\sigma_{0} = \sigma_1= 1$ and $P\rightarrow\infty$.

\subsection{Application to drug diffusion for a two-layer sphere}
We now consider the more general problem where the diffusivity varies in the concentric spheres. The following physical parameters are considered for computational experiments in both absorbing and desorbing cases for the core-shell spherical model \cite{hen}:
\begin{gather}
R_0=1.5 \cdot 10^{-3}\,\mathrm{m},  \qquad  R_1=1.7 \cdot 10^{-3}\, \mathrm{m},\qquad  \sigma_0=\sigma_1=1, \nonumber \\
\label{eq:params2}
 D_0=30 \cdot 10^{-11}\, \mathrm{m}^2 \mathrm{s}^{-1},  \qquad    D_1=5 \cdot 10^{-11}\, \mathrm{m}^2 \mathrm{s}^{-1},  \qquad  D_e=30 \cdot 10^{-11}\,  \mathrm{m}^2 \mathrm{s}^{-1},
\end{gather}
with $C_0=1$ and $C_e=0$ for the desorbing case and $C_0=0$ and $C_e=1$ for the absorbing case ($C_{\mathrm{max}}=1$ in both cases).
\bigskip
%

\begin{figure}[htbp]
\subfloat[$P\rightarrow\infty$ (uncoated)]{\includegraphics[width=0.5\textwidth]{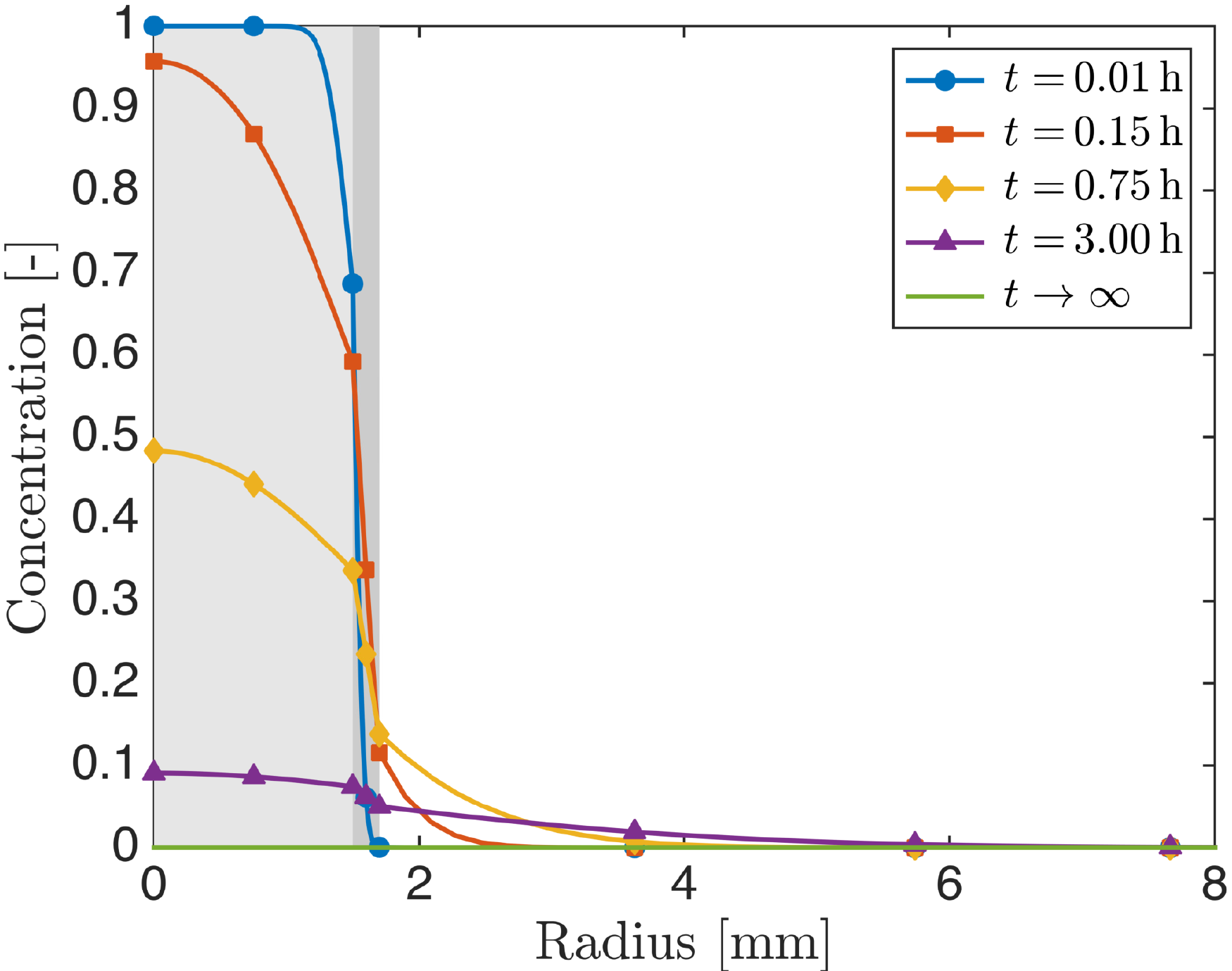}}\subfloat[$P=5\cdot10^{-8}$ (coated) ]{\includegraphics[width=0.5\textwidth]{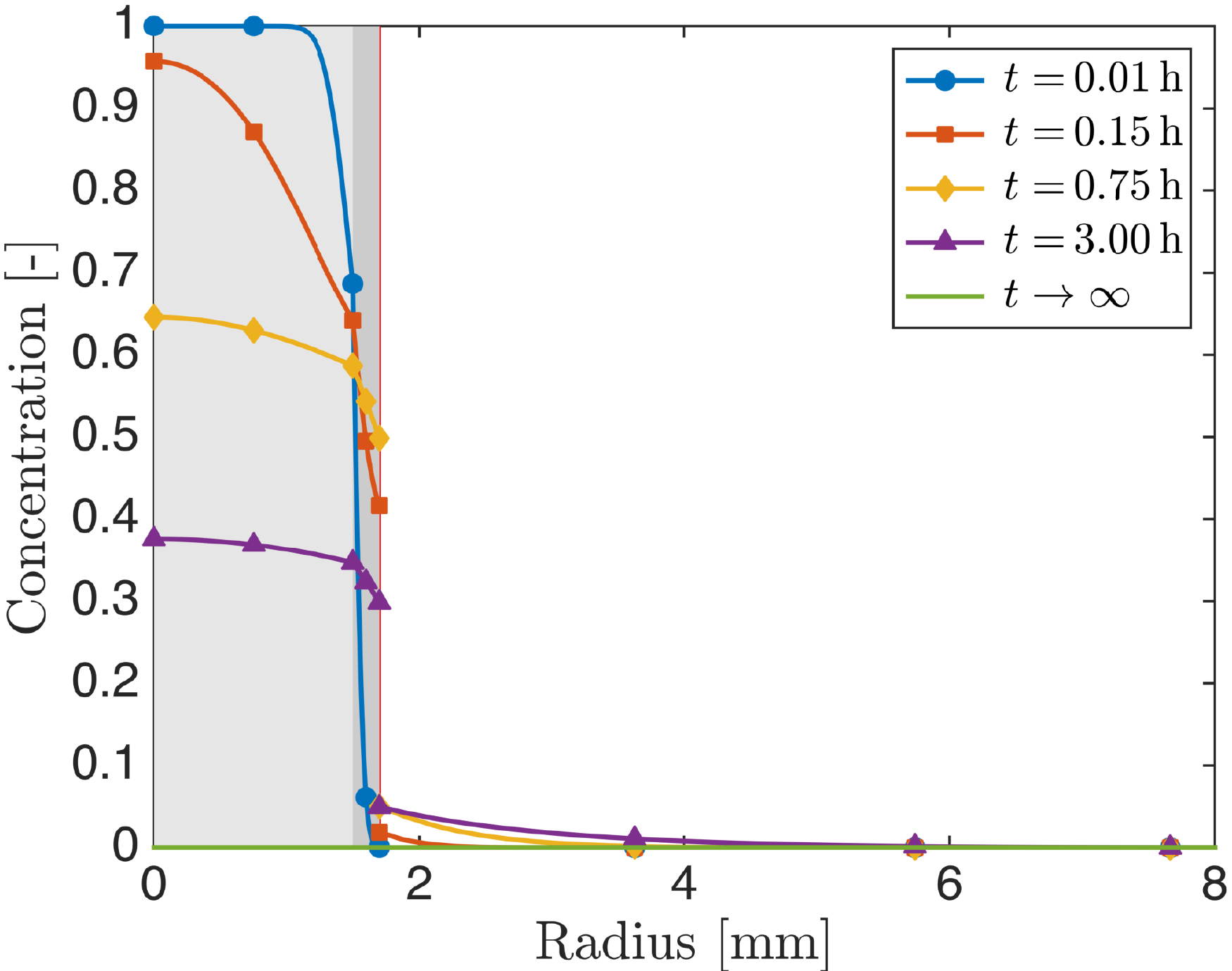}}\\
\subfloat[$P\rightarrow\infty$ (uncoated)]{\includegraphics[width=0.5\textwidth]{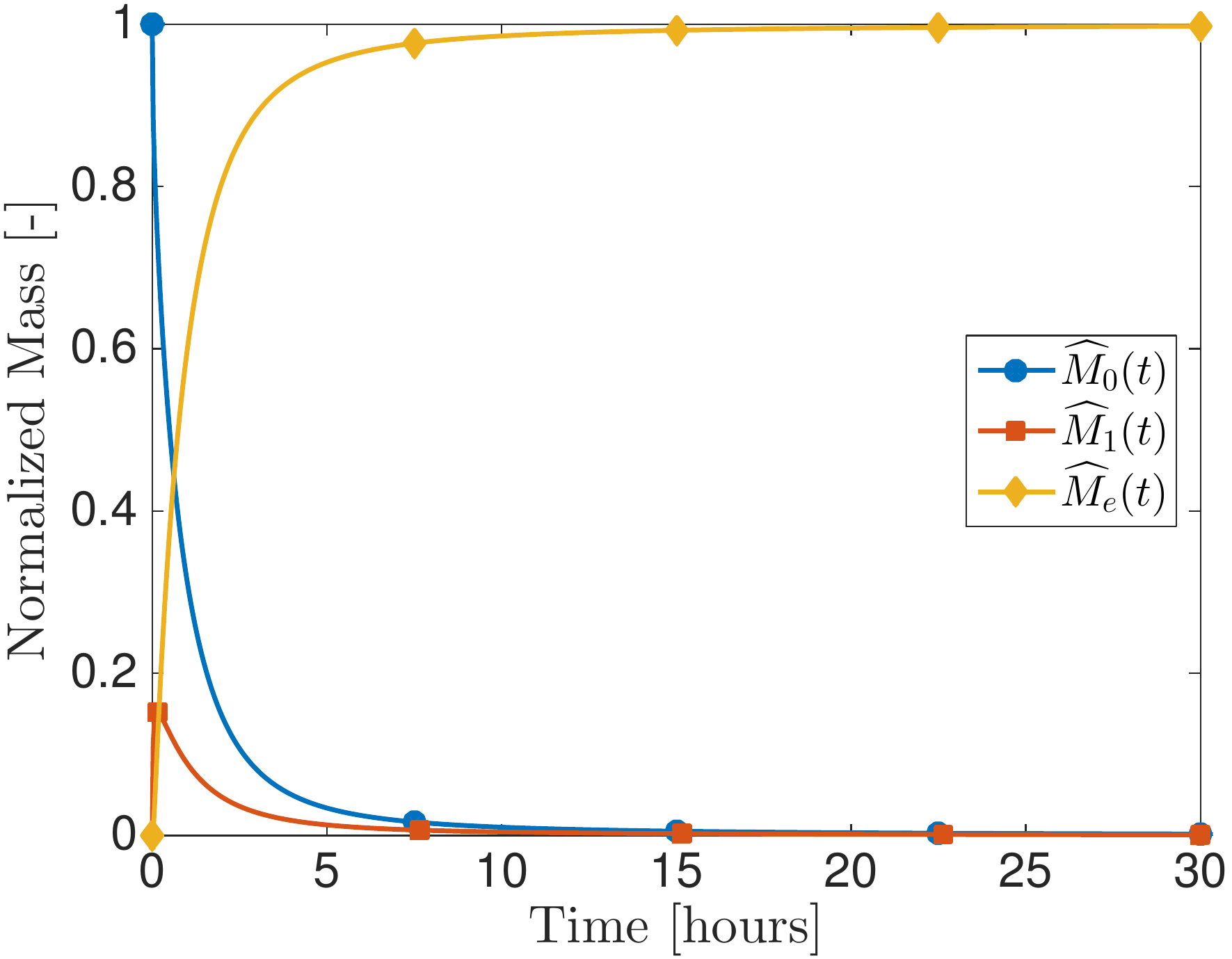}}\subfloat[$P=5\cdot10^{-8}$ (coated)]{\includegraphics[width=0.5\textwidth]{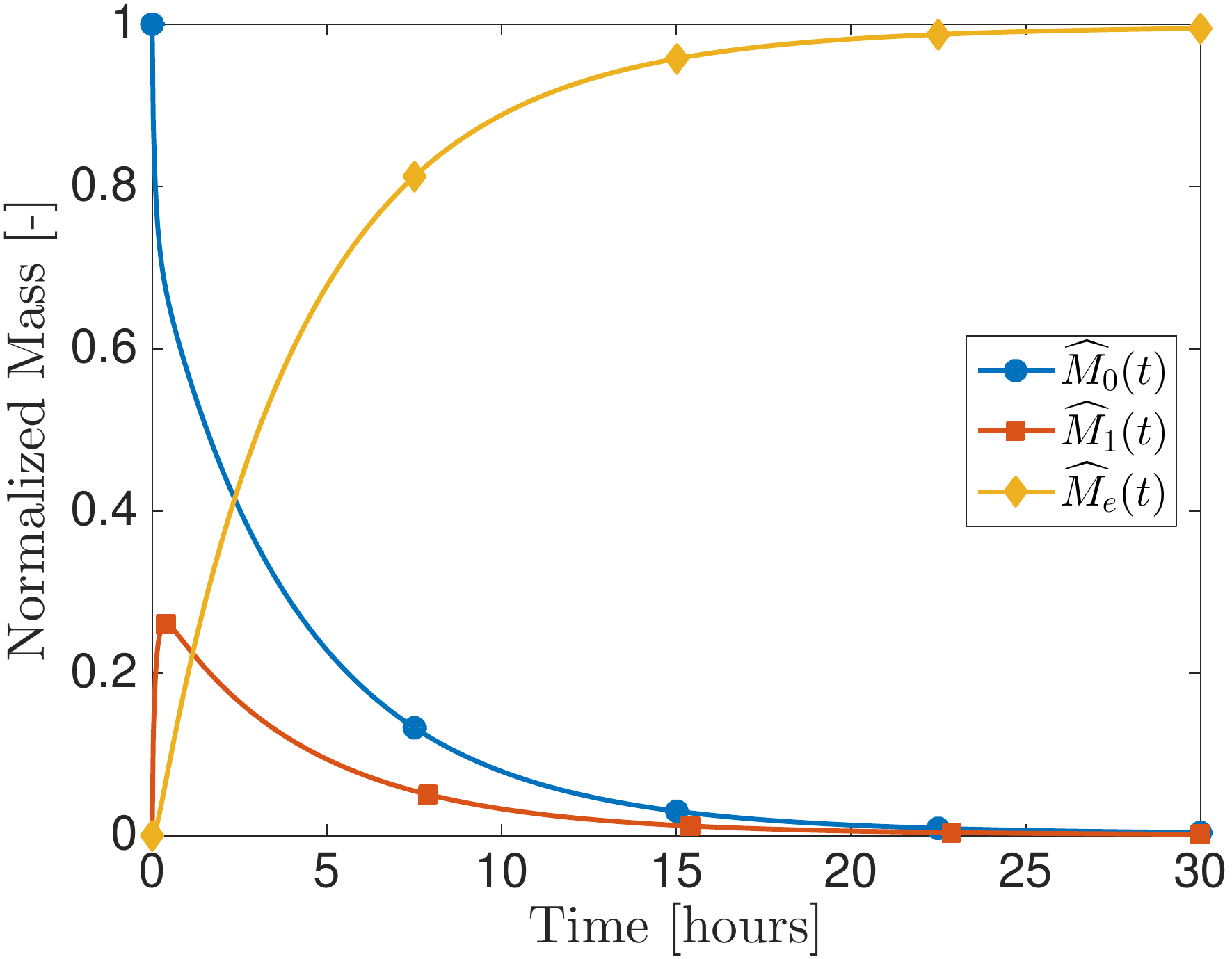}}
\caption{(a)--(b) Concentration profiles  for the two cases of an (a) uncoated and (b) coated desorbing microcapsule at several times. The core and shell layers are shaded in light and dark gray, respectively, while the thin coating shell is shaded in red (see Fig.~\ref{fig:1D_schematic}). The semi-infinite external medium is truncated at $r=8\,\mathrm{mm}$: beyond this point all concentrations remain constant. (c)--(d) Plot of the normalized drug mass, $\widehat{M}_{i}(t) = M_{i}(t)/(\frac{4}{3}\pi R_{0}^{3}C_{\mathrm{max}})$, over time in each layer for the two cases of an (c) uncoated and (d) coated desorbing microcapsule.}
\label{fig:desorbing}
\end{figure}

\noindent\underline{Desorbing case}\\
Among many applicative fields of releasing spheres, we focus here on layer-by-layer coated capsules, as controlled drug carriers. They have attracted significant attention for therapeutic applications and deserve special interest because of their potential for sustained release. For the desorbing case, the drug is transported from inner core, via the intermediate shell, to the release medium: each layer receives mass from the layer beneath it and transfers it to the layer above it, in a cascading sequence until the drug is completely released from the capsule. The coating mass transfer coefficient $P$ constitutes the distinctive parameter that controls the flux exiting the capsule. Fig.~\ref{fig:desorbing} shows the concentration profiles in the case of two different values of $P$:  $P \rightarrow \infty$ (uncoated sphere, Fig.~\ref{fig:desorbing}a) and $P = 5\cdot10^{-8}$ (coated sphere, Fig.~\ref{fig:desorbing}b). Concentration is decreasing inside each layer and is discontinuous at the interlayer interfaces for finite $P$, with the mass flux continuity preserved (Fig.~\ref{fig:desorbing}ab). Excellent agreement is achieved when comparing our results to those presented in \cite{pon}.

\begin{figure}[htbp]
\centering
\subfloat[$P=10^{-3}$]{\includegraphics[width=0.5\textwidth]{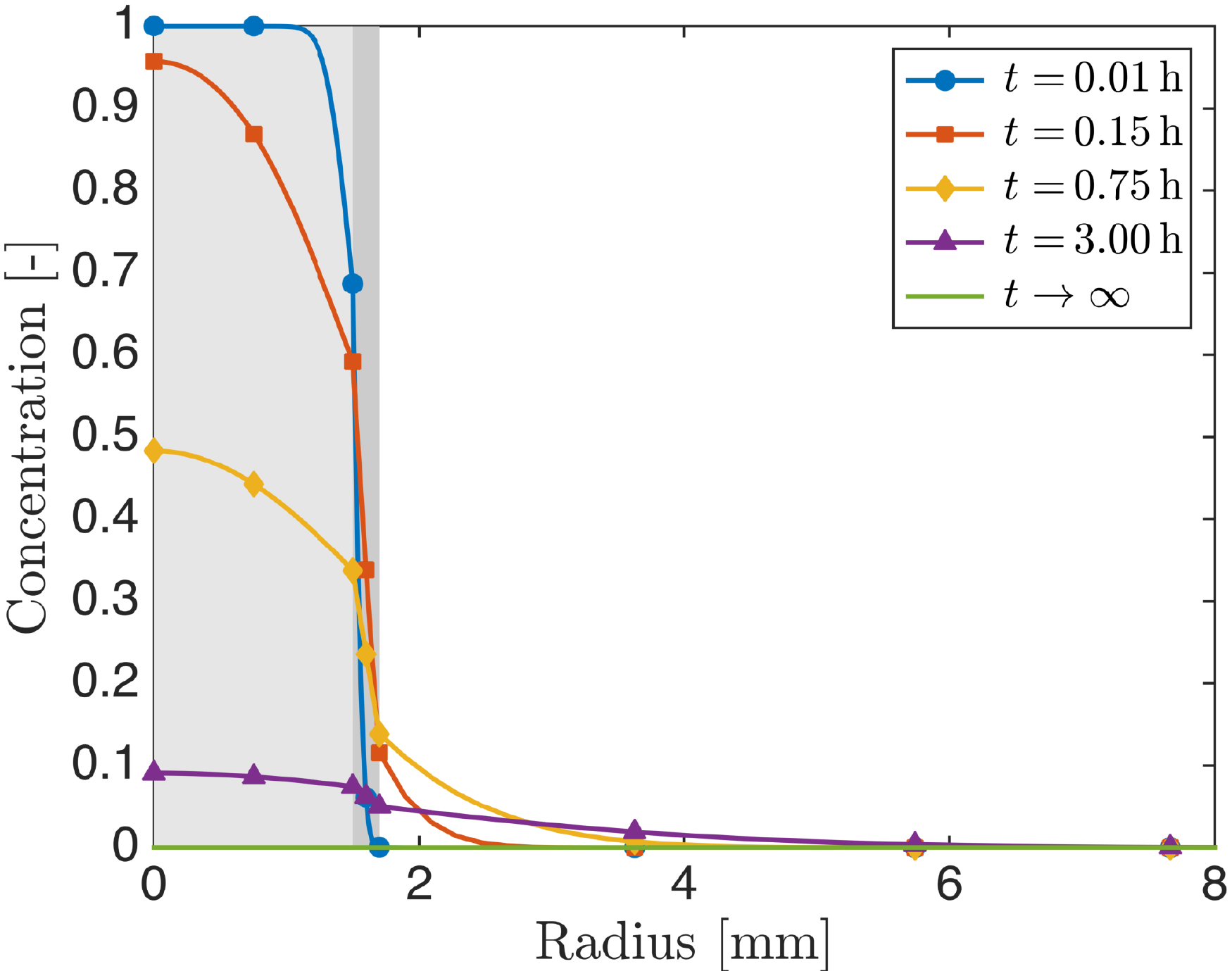}}
\subfloat[$P=10^{-8}$]{\includegraphics[width=0.5\textwidth]{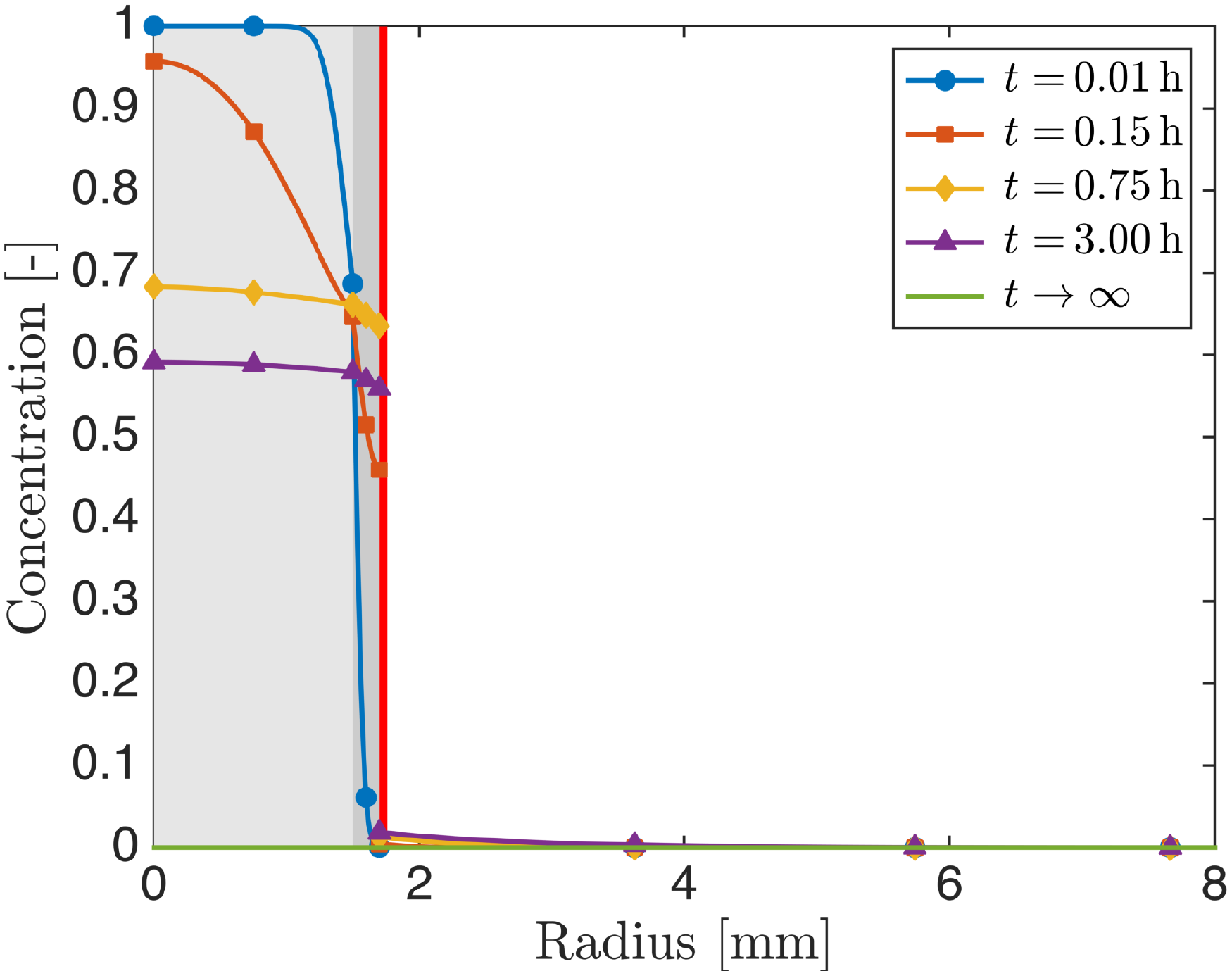}}\\
\subfloat[$P=10^{-3}$]{\includegraphics[width=0.5\textwidth]{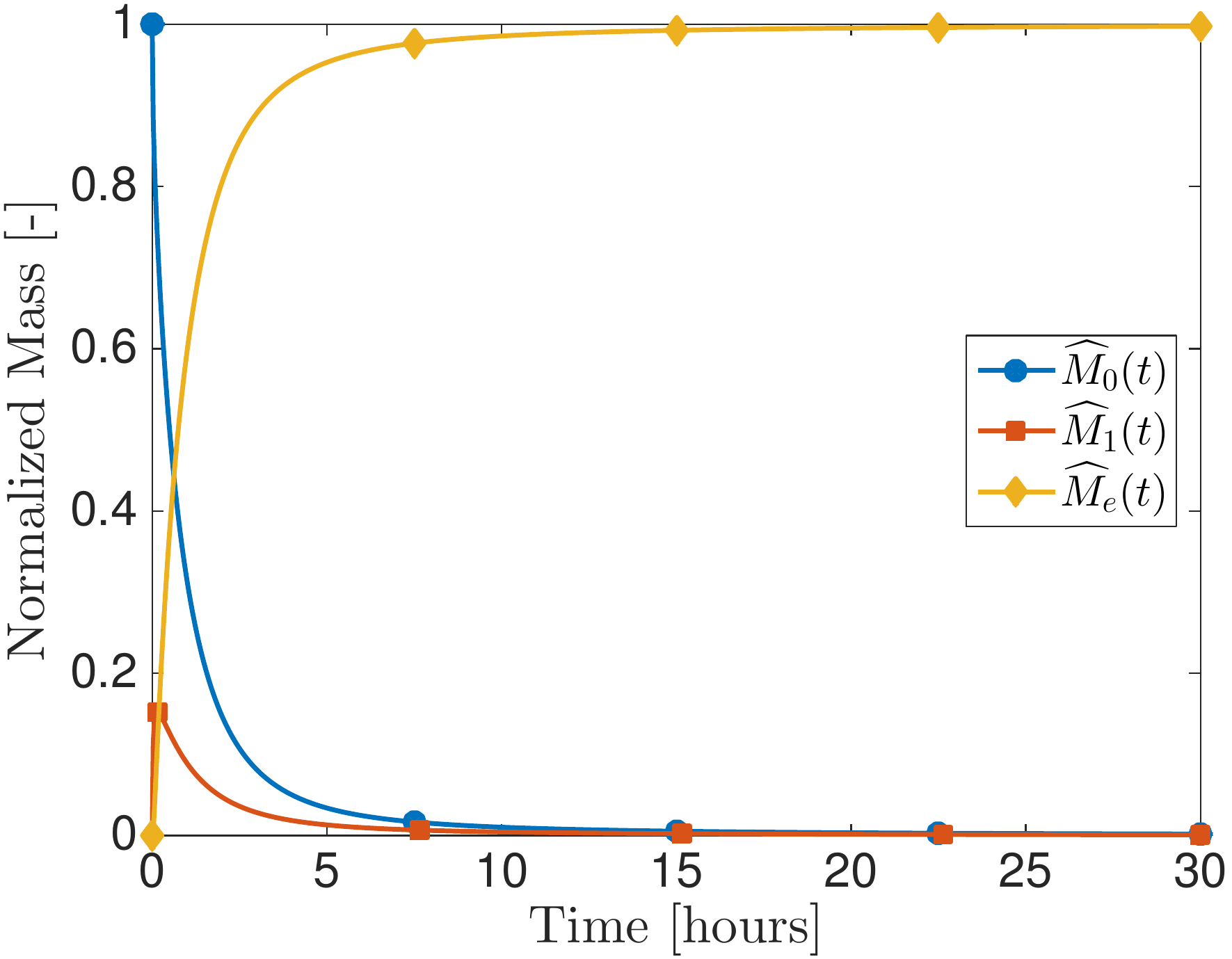}}
\subfloat[$P=10^{-8}$]{\includegraphics[width=0.5\textwidth]{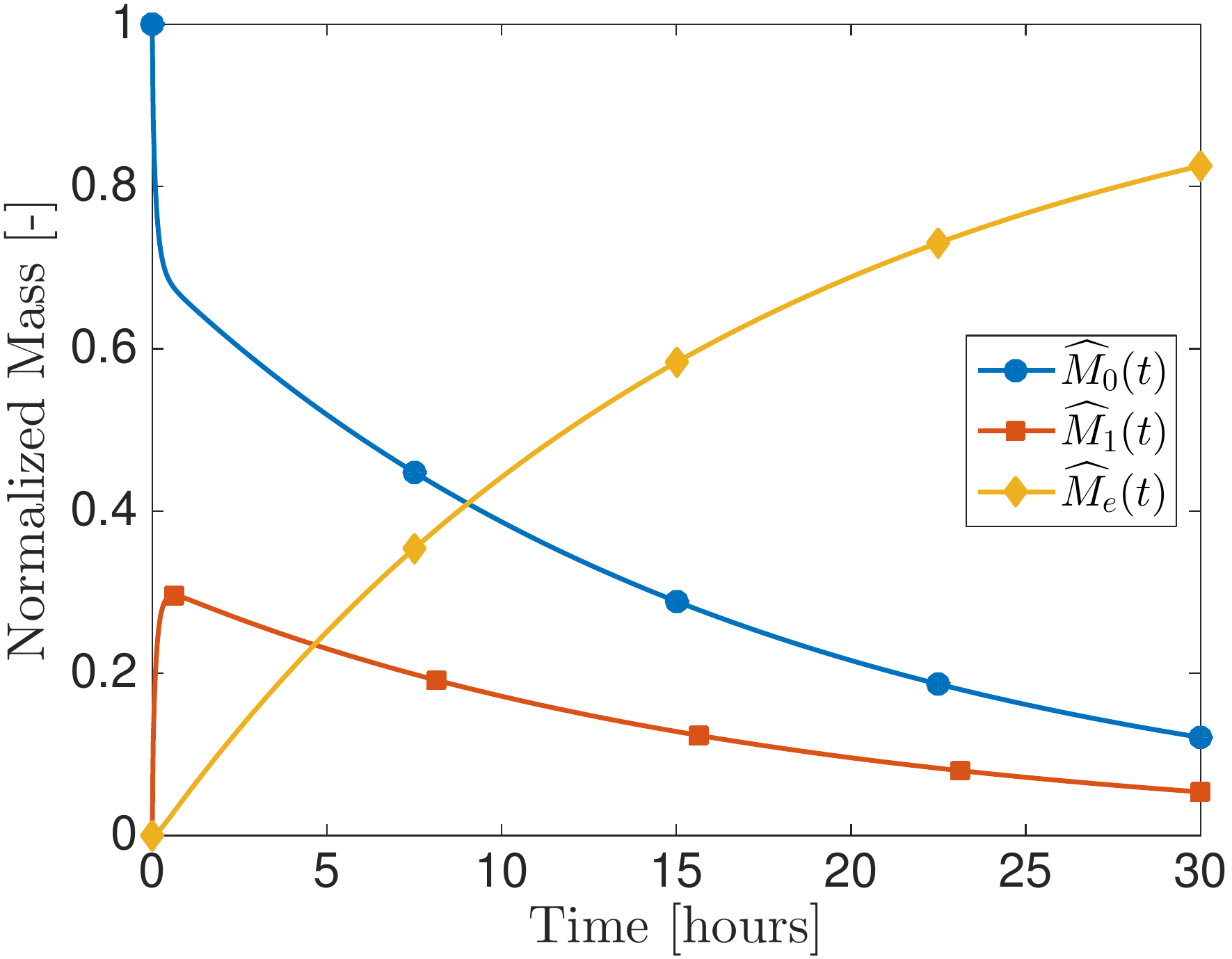}}
\caption{Effect of mass transfer coefficient at coating layer for the desorbing case. (a)--(b) Concentration profiles for two values of $P$ at the same times as those shown in Fig.~\ref{fig:desorbing}. The core and shell layers of the microcapsule are shaded in light and dark gray, respectively, while the thin coating shell is shaded in red (see Fig.~\ref{fig:1D_schematic}). The semi-infinite external medium is truncated at $r=8\,\mathrm{mm}$: beyond this point all concentrations remain constant. (c)--(d) Plot of the normalized mass, $\widehat{M}_{i}(t) = M_{i}(t)/(\frac{4}{3}\pi R_{0}^{3}C_{\mathrm{max}})$, over time in each layer for two values of $P$.}
\label{fig:P_effect}
\end{figure}

Due to the sink boundary condition (\ref{erf0}), all mass eventually accumulates in the external release medium. In other words, due to the condition (\ref{erf0}), all drug mass is released into the environment after a sufficiently long time and the total mass is preserved. The drug mass monotonically decreases in the core (layer 0), while at the same time increasing up to some peak before decaying asymptotically in the hydrogel layer (layer 1) (Fig.~\ref{fig:desorbing}cd, with the mass normalized by its initial value $\frac{4}{3}\pi R_0^3 C_{\mathrm{max}}$). In the release medium, the mass progressively increases at a rate depending on the diffusive properties of the two-layer materials. The simulation indicates that the time and the size of the mass peak in the hydrogel layer (layer 1) is related to the releasing properties of the core, on the one hand, and to the diffusivity of the release medium, on the other hand, together with the mass resistance of the coating. The thin hydrogel layer retains a negligible mass due to its thickness, and the core is completely emptied after roughly 10 hours, in the case of $P\rightarrow\infty$. After that time, all the mass is transmitted to the external medium. A much more sustained release occurs in the case of a coating having a finite and small mass transfer coefficient ($P = 5\cdot10^{-8}$). After 10 hours, a substantial amount of drug remains in the core and hydrogel layers, with the core not completely empty until approximately 22.5 hours.

\begin{figure}[!t]
\subfloat[$P\rightarrow\infty$ (uncoated)]{\includegraphics[width=0.5\textwidth]{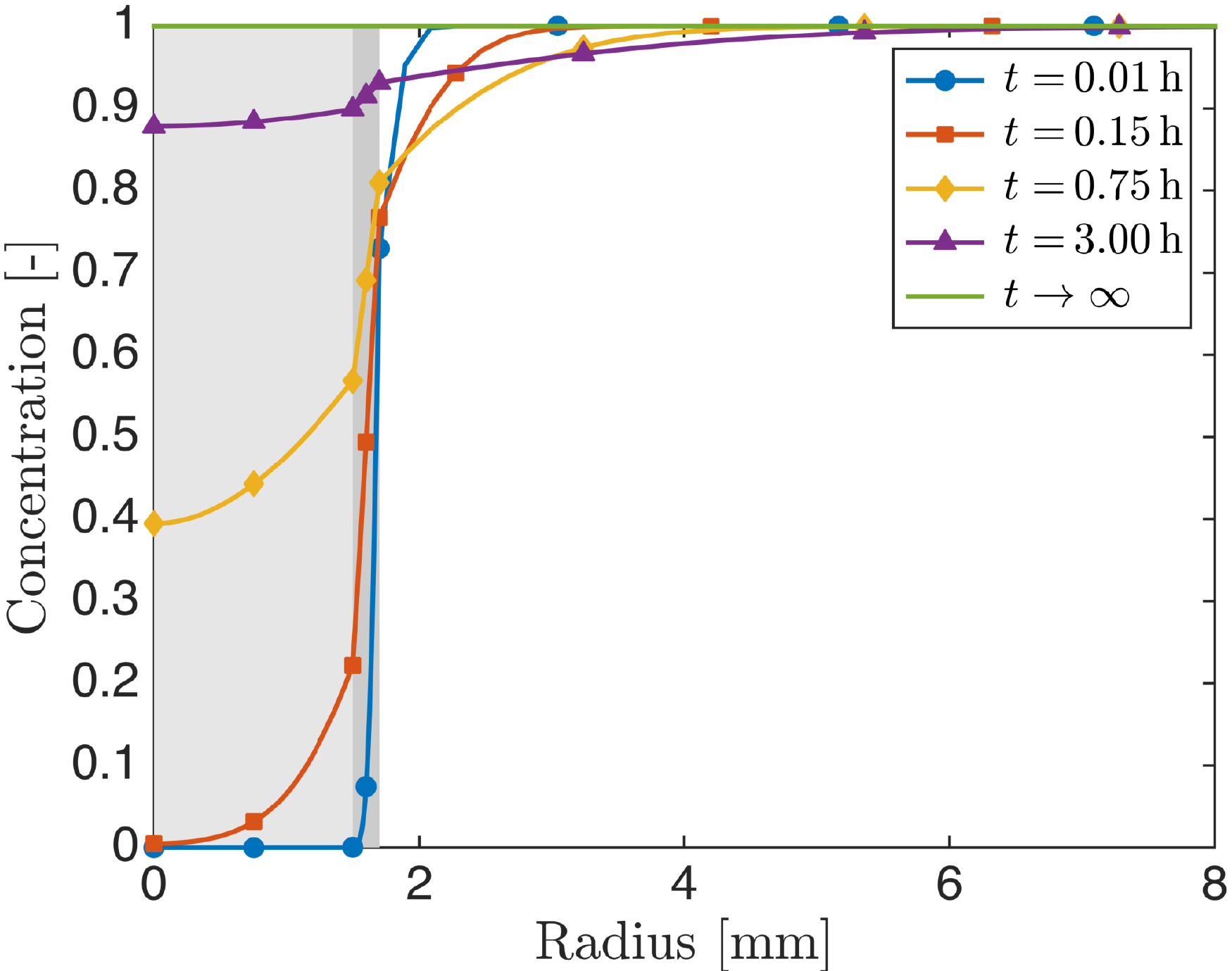}}
\subfloat[ $P=5\cdot10^{-8}$ (coated) ]{\includegraphics[width=0.5\textwidth]{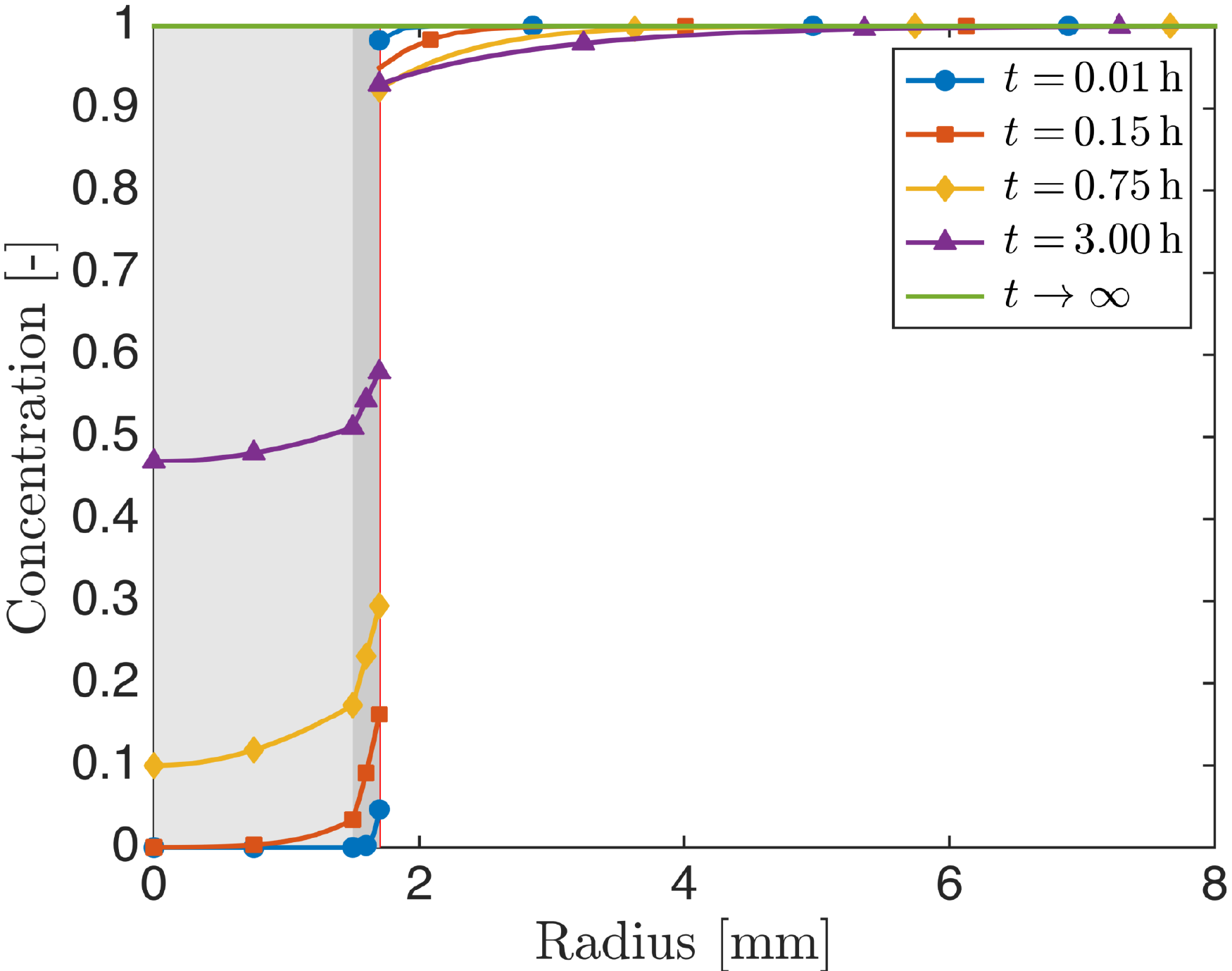}}\\
\subfloat[$P\rightarrow\infty$ (uncoated) ]{\includegraphics[width=0.5\textwidth]{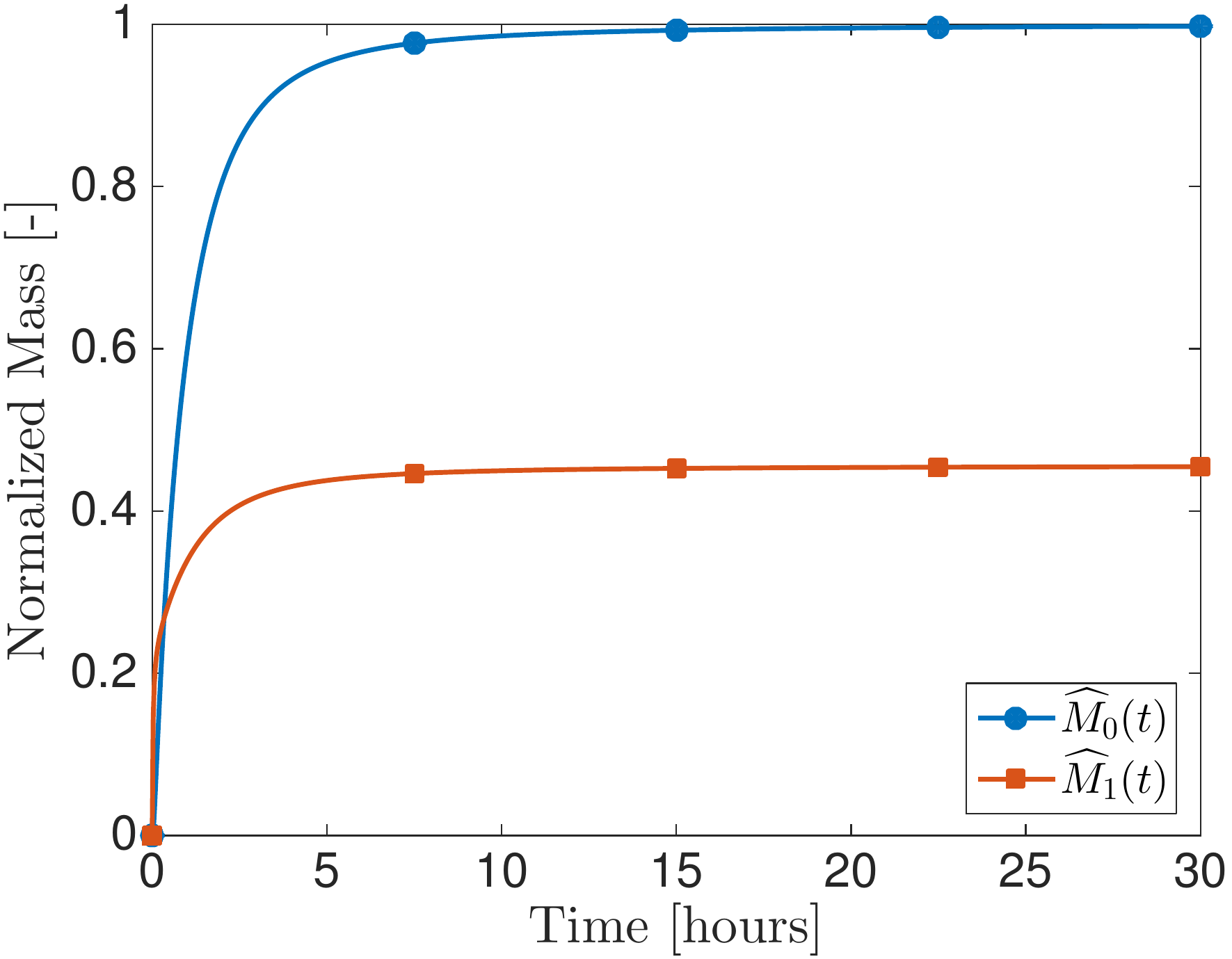}}
\subfloat[$P=5\cdot10^{-8}$ (coated) ]{\includegraphics[width=0.5\textwidth]{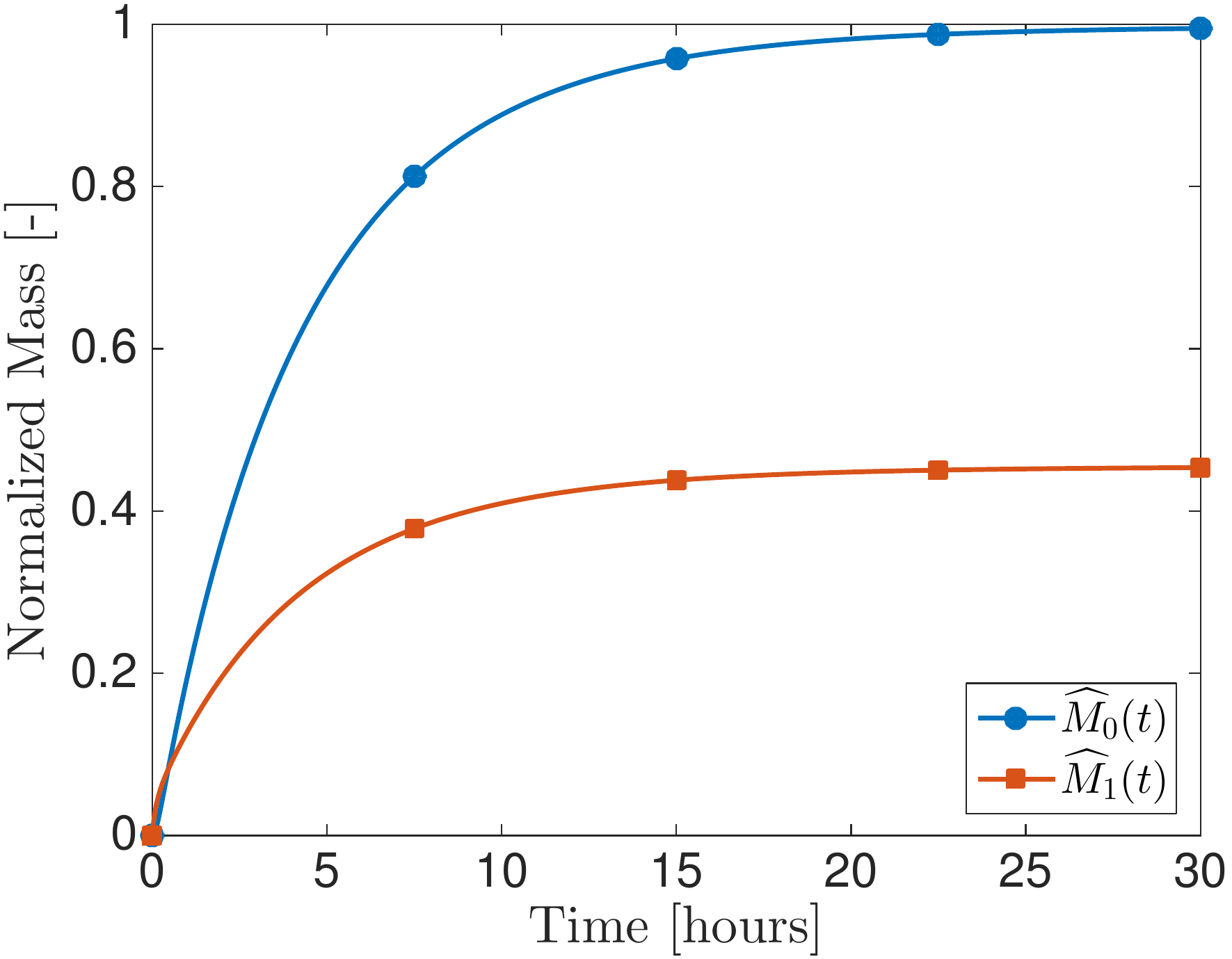}}
\caption{(a)--(b) Concentration profiles  for the two cases of an (a) uncoated and (b) coated absorbing microcapsule at times as in figs. 4 and 5. The semi-infinite external medium is truncated at $r=8\,\mathrm{mm}$: beyond this point all concentrations remain constant. The core and shell layers  are shaded in light and dark gray, respectively, while the thin coating shell is shaded in red (see Fig.~\ref{fig:1D_schematic}). (c)--(d) Plot of the normalized drug mass, $\widehat{M}_{i}(t) = M_{i}(t)/(\frac{4}{3}\pi R_{0}^{3}C_{\mathrm{max}})$, over time in each layer  for the two cases of an (c) uncoated and (d) coated absorbing microcapsule. The normalized mass in the external medium is not shown since it is infinite for all time.}
\label{fig:absorbing}
\end{figure}

We now investigate the sensitivity of the solution to the value of coating mass transfer coefficient $P$ as an effective rate-controlling parameter. It turns out that for the above parameters (\ref{eq:params2}), the sensitive values of $P$ are in the range: $10^{-8} \leq P \leq 10^{-3}$. For $P=10^{-8},$ the coating almost acts as an impermeable barrier with a very small transfer rate from the capsule to the external medium evident in Fig. \ref{fig:P_effect}b. Interestingly, decreasing $P$ by a factor of 5 from the case $P = 5\cdot10^{-8}$ considered earlier (Figs.~\ref{fig:desorbing}bd) has a huge effect on the release rate: after 30 hours roughly 20\% of the mass still remains in the capsule (core and hydrogel layers, Fig.~\ref{fig:P_effect}bd). Setting $P=10^{-3}$ produces results that are indistinguishable from $P\rightarrow\infty$ as there is no observable difference between Figs \ref{fig:desorbing}ac and \ref{fig:P_effect}ac. In this case, the capsule surface is in perfect contact with the external ambient medium as evident by the continuity in concentration (Fig.~\ref{fig:P_effect}a).\\

\noindent\underline{Absorbing case }\\
The process of drug absorption from a saturated solution determines, in part, its bioavailability and is the basis of \textit{in-vitro} experiments. Drug kinetics from environment into an absorbing sphere is very similar to the desorbing case, except that the initial mass is present in the external semi-infinite medium and the drug transport direction is reversed. Fig.~\ref{fig:absorbing} (with mass normalized by $\frac{4}{3}\pi R_0^3 C_{\mathrm{max}}$) is the counterpart of Fig.~\ref{fig:desorbing} and shows how, and to what extent, the drug diffuses into the two-layer sphere. In the case of a coated sphere, the diffusion rate is lower and the drug reaches saturation after a longer period of time. In contrast to the desorbing case, the polymeric shell (layer 1) fills up to a maximum concentration and receives diffused mass from the external source, taken as a large reservoir, and transfers it to the inner core. The normalized saturation mass in layer 1 (Fig.~\ref{fig:absorbing}cd) depends only on the geometrical configuration (see Eq.~(\ref{mass_abs})): 
\begin{gather*}
\lim_{t\rightarrow\infty} \,\frac{M_{1}(t)}{\frac{4}{3}\pi R_{0}^{3}C_{\mathrm{max}}} = \frac{R_1^3 -R_0^3}{R_0^3} = \left(\frac{R_1}{R_0} \right)^3 -1 \simeq 0.45.
\end{gather*}

\section{Conclusions}
 
A better understanding of the mass transfer from a drug carrier or a vehicle in a living tissue for therapeutic purposes constitutes an important challenge in medicine nowadays. Mathematical modelling helps in predicting the drug release rates and diffusion behavior from these delivery systems, thereby reducing the number of experiments needed.\par
In the current work, inspired by the above biomedical application, we have presented a mathematical model and developed an analytical technique to study the diffusion-controlled mass desorption-absorption systems in microspheres. We have focussed on drug release from two-layer spherical capsules, consisting of an inner core and an outer shell protected by a thin coating, immersed in an external semi-infinite medium. This type of geometry, introducing additional design parameters to the formulation (i.e. relative sizes and relative drug diffusion and partition coefficients of inner and outer structures), enriches the possibilities in terms of pharmacokinetics, controlled release rate and, ultimately, delivery performance. As in many biological systems, the model contains a number of parameters, subject sometimes to high variability and uncertainty, that need to be identified before it can be used in a predictive way to provide the drug kinetics. Once the parameters are identified, the proposed methodology provides a simple tool that can be used to quantitatively characterize the drug diffusion, improve the technological performance and optimize the release rate for therapeutic purposes. By virtue of the one-to-one analogy of mass diffusion and heat conduction problems, the presented approach can be successfully applied to the similar model of heat transfer from/in a sphere immersed in a large ambient medium.

\section*{Acknowledgments}
The first author acknowledges support from the Australian Research Council (DE150101137).

\end{document}